\documentclass[a4paper,11pt]{article}
\pdfoutput=1

\usepackage{jheppub}
\usepackage{graphics, setspace}

\newcommand{\mathsym}[1]{{}}
\newcommand{\unicode}[1]{{}}

\usepackage{color}
\usepackage{graphicx}                   
\usepackage{bbm}
\usepackage{amsmath}
\usepackage{amssymb}
\usepackage{bbold}
\usepackage{xspace}
\usepackage{verbatim}
\usepackage{bm}

\newcommand{\nbar}{{\bar n}}

\DeclareRobustCommand{\Eq}[1]{Eq.~(\ref{#1})}

\DeclareRobustCommand{\Ref}[1]{Ref.~\cite{#1}}

\usepackage{color}
\definecolor{darkblue}{rgb}{0,0,0.5}
\definecolor{darkred}{rgb}{0.5,0,0}
\definecolor{orange}{rgb}{0.6,0.3,0.1}
\definecolor{purple}{rgb}{0.5,0.0,0.5}

\newcommand{\Gnorm}{\,\!}

\DeclareRobustCommand{\Eq}[1]{Eq.~(\ref{#1})}

\DeclareRobustCommand{\Ref}[1]{Ref.~\cite{#1}}

\begin{document}

LA-UR-18-22121

\title{ Soft Evolution After a Hard Scattering Process}

\author{Duff Neill,}
 \author{Varun Vaidya}%
\affiliation{%
Theoretical Division, MS-248, Los Alamos National Laboratory, Los Alamos, NM 87545
}%

\emailAdd{duff.neill@gmail.com}
\emailAdd{vvaidya@lanl.gov}

\abstract{The dynamical cascade of momentum, spin, charge, and other quantum numbers from an ultra-violet process into the infra-red is a fundamental concern for asymptotically free or conformal gauge field theories. It is also a practical concern for any high energy scattering experiment with energies above tens of GeV. We present a formulation of the evolution equation that governs this cascade, the Banfi-Marchesini-Smye equation, from both an effective field theory point of view and a direct diagrammatic argument. The equation uses exact momentum conservation, and is applicable to both scattering with initial and final state hard partons. The direct diagrammatic formulation is organized by constructing a generating functional. This functional is also automatically realized with soft wilson lines and collinear field operators coupled to external currents. The two approaches are directly connected by reverse engineering the Lehman-Symanzik-Zimmermann reduction procedure to insert states within the soft and collinear matrix elements. At leading order, the cascade is completely controlled by the soft anomalous dimension. By decomposing the anomalous dimension into on-shell and off-shell regions as would be realized in the effective field theory approach with a Glauber mediating potential, we are forced to choose a transverse momentum ordering in order to trivialize the overlap between Glauber potential contributions and the pure soft region. The evolution equation then naturally incorporates factorization violating effects driven by off-shell exchanges for active partons. Finally, we examine the consequences of abandoning exact momentum conservation as well as terminating the evolution at the largest inclusive scale, procedures often used to simplify the analysis of the cascade.   }

\maketitle

\section{Introduction}

We are concerned with the dynamics of the reduced density matrix for soft or collinear physics at late times after a hard interaction. This has been discussed extensively in Refs. \cite{Nagy:2007ty,Nagy:2008eq,Nagy:2008ns,Nagy:2012bt,Nagy:2014mqa,Nagy:2017ggp}, with the aim of developing a shower which coherently simulates the transport of color and spin into the infra-red, including simultaneously soft and collinear effects. We will examine a simplified evolution of the reduced density matrix, focusing mostly on soft physics, since our aim is not to replicate previous efforts.\footnote{Though as will be pointed out in Sec. \ref{eq:Beyond_Everything}, even in soft-sensitive observables, collinear effects can enhance certain regions of phase-space \cite{Neill:2015nya,Larkoski:2016zzc}, destabilize the soft approximation for the evolution kernel \cite{Iancu:2015vea,Iancu:2015joa,Hatta:2017fwr}, or even dominate certain limits of the distribution as found in Ref. \cite{Neill:2016stq}.} We focus on the soft approximation since non-trivial characteristics of the QCD parton-shower can still be obtained. For instance the structure of the color charge transport is one of the most questions, being connected to the violation of collinear factorization when both initial and final hard directions are present in the scattering\cite{Catani:1985xt,Forshaw:2006fk,Forshaw:2008cq,Forshaw:2012bi,Catani:2011st,Angeles-Martinez:2015rna,Schwartz:2017nmr}. Nonetheless, the evolution and factorization of the density matrix is well-defined, with or without this additional collinear factorization. Further, the soft approximation manifests the duality of the time-like evolution of the parton shower and the space-like evolution of small-x physics (Refs. \cite{Balitsky:1995ub, Kovchegov:1999yj,JalilianMarian:1996xn,JalilianMarian:1997gr,Iancu:2001ad}) found in Refs. \cite{Marchesini:2003nh,Hatta:2008st,Avsar:2009yb}.The soft approximation also simplifies the writing of the evolution equations for the parton shower, which is critical to our chief concern: to know what logs are resummed in any parton shower given the definition of the shower, what are the consequences of working in certain simplifying approximations, and what are the dynamical structures developed in the course of the cascade.  Finally, we would like to know how to map the shower construction to matrix elements of soft wilson lines and collinear operators, of the sort used soft-collinear effective field theory (SCET) \cite{Bauer:2000yr,Bauer:2000ew,Bauer:2001ct,Bauer:2001yt,Bauer:2002nz}. Our aim is to provide tools to aid these questions.

We will present a simple way to organize the evolution of the density matrix:\footnote{The astute would note that given the markovian nature of the QCD parton shower in the leading approximation, the reduced density matrix should obey a Lindblad equation on very general grounds of quantum mechanics, see Refs. \cite{KOSSAKOWSKI1972247,Lindblad:1975ef,Gorini:1976cm}, if we accept that we are ordering in momentum space rather than in time. In the context of the soft evolution, it does, and this equation is nothing other than the Banfi-Marchesini-Smye (BMS) equation \cite{Marchesini:2004ne}.}  one using generating functionals to automate the recursive insertion of real and virtual soft gluons, rather than the prescriptive rules of Refs. \cite{Bassetto:1982ma,Bassetto:1984ik,Catani:1984dp,Catani:1985xt,Fiorani:1988by,Berends:1987me,Berends:1988zn,Mangano:1990by}. Though this generating functional seems like a diagrammatic construction, we will also show that it is nothing other than a particular matrix element of soft wilson lines and collinear field operators, directly connected to the ``jet substructure'' factorization properties in Ref. \cite{Larkoski:2015zka}. The two approaches are connected by using some formal manipulations of the Lehman-Symanzik-Zimmerman (LSZ) reduction formula. These soft wilson lines are objects familiar to the effective field theorist, found naturally in the context of soft-collinear effective field theory. Indeed, using a rather effective field theory line of reasoning based on recent developments of SCET with off-shell potentials, Ref. \cite{Rothstein:2016bsq}, we will conclude the parton shower must be ordered using the transverse momentum of the next splitting.

A critical question which we will consider is the use of the multi-pole expansion of effective theories in defining the evolution equations of the reduced density matrix. Loosely speaking, the multipole expansion simplifies recoil: when a particular variable is multipole expanded, the two sectors no longer feel the momentum recoil of each other. For example, if we are given light-cone directions $n,\bar{n}$, $n\cdot\bar{n}=2$, and a local product of collinear and soft operators $O_{n}(x)$ and $O_s(x)$ defined on momenta with scaling:
\begin{align}
 p &=(\bar{n} \cdot p,n \cdot p,p_{\perp})\,,\\
p_n&= Q(1,\lambda^2,\lambda)\,,\\
p_s&= Q(\lambda,\lambda,\lambda)\,.
\end{align}
Where we have introduced some small parameter $\lambda\ll 1$ and a hard momentum scale $Q$. Then we would expand the local product as:
\begin{align}
O_n(n\cdot x,\bar{n}\cdot x,x_\perp)O_s(n\cdot x,\nbar\cdot x,x_\perp)&=O_n(n\cdot x,0,x_\perp)O_s(0,\nbar\cdot x,x_\perp)+...
\end{align}
That is, since the $n\cdot p$ component of the soft operator is larger than the collinear, and the $\bar{n}\cdot p$ momentum component of the collinear operator is larger than the soft, the expansion will now prevent the soft momenta from participating in the conservation of the $\bar{n}$-components, and the collinear from participating in the conservation of the $n$-components. Both operators will participate in the conservation of the transverse momenta. This can be seen examining the fourier transforms of the operators, after the expansion. To a leading power approximation, momentum conversation will be respected none-the-less when summing over all momentum sectors that can contribute to the observable. 

The multipole expansion as initially formulated in Ref. \cite{Grinstein:1997gv} is a means to ensure that one has homogeneous power counting in the expansion at each order in the effective theory, and is now a standard tool applied to effective theory constructions. Usually applying the multipole expansion on operators of QCD fields will result in additional divergences that facilitate the resummation of logarithms, while also removing overlap between distinct EFT sectors. Most parton showers, on the other hand, are not conducted with any sort of multipole expansion\footnote{Two exceptions are Refs. \cite{Dasgupta:2001sh,Dasgupta:2014yra}, though those papers never use the language of ``a multipole expansion'' in defining their algorithms. They \emph{do} intentionally ignore momentum conservation for the exact purpose of isolating \emph{only} the soft or the \emph{only} collinear contributions to the observables being resummed, which is part of the purpose of the multipole expansion.}. They instead maintain momentum conservation at each step in the shower. When coupled with an ordering variable, this will regulate all infra-red divergences at each splitting stage in the shower. 

When formulating our shower equations with regards to momentum conservation, we will follow the parton shower approach. This is perhaps a heretical position for avowed effective field theorists, but we have a sound reason for doing so. Not only would this be necessary for a generic parton shower that could be applied to any observable (since the power counting necessary for the multipole expansion would never be known a priori), we in addition will argue that following a naive multipole expansion with an insufficient number of EFT sectors induces a renormalon like effects in the parton shower \cite{Larkoski:2016zzc}, based on whether or not one keeps jet functions in the factorization. Disregarding the collinear contributions still results in a perfectly consistent EFT, but one suffering from large logs in the phase-space being integrated over. This can be seen by examining the low scale matrix elements found in a multipole expanded EFT set-up for hemisphere jet observables in $e^+e^-$, as in Refs. \cite{Becher:2016mmh,Becher:2016omr}, where the jet function contribution is indeed dropped. Of course, one can power count the collinear sectors differently, keeping certain jet functions, as in Ref. \cite{Neill:2015nya}, or we can just never perform the multipole expansion, allowing the exact phase space constraints of the measurement to correctly constrain all multiple emissions, and evolve our parton shower down to \emph{zero} cutoff for the ordering variable, rather than the maximal momentum scale below which we are completely inclusive. Following exact momentum conservation is much closer to in spirit the automated approach to generic next-to-leading logarithm resummation approach found in the CAESAR program (Ref. \cite{Banfi:2004yd}).\footnote{In practice of course, we would not evolve down to zero ordering variable, but to a scale just above the breakdown of perturbation theory, matching to truly non-perturbative evolution there-after.}

{\bf Note:} As this paper was being finished, Ref. \cite{Martinez:2018ffw} appeared, which also considered the full-color evolution of the parton shower with both initial and final state hard partons. Importantly, the paper addresses how one can numerically implement the full-color evolution, and gives approximations that perturb around the large $N_c$ limit, see also Refs. \cite{Platzer:2012np,Platzer:2013fha}. 

\section{Outline of Paper}
The logic of this paper is as follows: we first introduce the concept of the reduced density matrix, first formulated in terms S-matrix elements, but then also in terms of so-called N-jet operators used in the SCET factorization of amplitudes. We briefly describe how one calculates ``late-time'' observables with this object. We then show how one can, from the bottom up, formulate a generating functional that summarizes all possible soft real emissions from a particular hard scattering, similar to Ref. \cite{Weigert:2003mm}. This generating functional is explicitly worked out to leading order in real and virtual emissions. To go to higher orders, one simply needs to calculate soft currents to whatever order in perturbation theory one can, while dressing the hard scattering with the appropriate virtual corrections. The utility of the generating functional is in how it summarizes the eikonal factorization of soft partons: soft gluon insertions are simply matter of defining the correct derivative operator, which is completely dictated by the soft currents. Inserting soft gluons amounts to taking derivatives. After we familiarize ourselves with this generating functional technique by working out the ``leading-log'' generating functional, where all soft emissions are given by the tree-level eikonal feynman rule. We then square these generating functionals, and using soft gluon insertion operators working at the level of the amplitude squared, show how one can construct the Banfi-Marchesini-Smye (BMS) equation \cite{Marchesini:2004ne}. Solving the BMS equation amounts to resumming the trace one would take of the reduced density matrix weighted by the appropriate observable. We also show the necessity of using the transverse momentum of the soft gluons as the ordering variable and discuss how off-shell exchanges involving active partons are included (however, our proof rests on a decidedly effective field theory arguement). As an illustration of the resummation of large logarithms accomplished by the BMS equation, we recover the CAESAR (Ref. \cite{Banfi:2004yd}) resummation formula for global logarithms by examining the full evolution equation in a particular kinematic limit.\footnote{Technically speaking, we only recover the double-logarithmic terms in the CAESAR formula, but if we trivially promoted our eikonal functions to the correct antennae or Catani-Seymour subtraction functions, we would capture all single logarithmic contributions, including any non-global contributions.} This application is particularly fascinating, since the momentum regions usually assigned in the SCET power counting naturally arise when using the full momentum conservation and transverse ordering of the shower.

Then we turn our attention to effective field theory objects like matrix elements of soft wilson lines. Typically these are an observable weighted sum over final state soft emissions, with the soft states acting on time-ordered and anti-time-ordered wilson lines given by the paths taken by energetic particles. We show how one can rewrite the trace over the soft final state in terms of functional derivatives acting on external currents probing the wilson lines, which amounts to the LSZ reduction procedure for S-matrix elements. For certain classes of observables, this LSZ operator implementing the insertion of a complete set of states constrained by a measurement can be written recursively in terms of an integral equation. Then using the so-called SCET$_+$ factorization, we argue that the factorization properties of these wilson lines are exactly those of the generating functionals we earlier constructed, ultimately pointing to the equivalence of the two approaches. Given that the resummation properties of soft wilson lines are organized by the soft anomalous dimension, this then justifies using the effective field theory reasoning to conclude the transverse ordering of the shower in the BMS equation. We end the SCET section by giving a derivation of the BMS equation in the hemisphere jets case, now using a naive power counting of only hard and soft sectors, corresponding to the emissions in the energetic hemisphere, and generic inclusive soft radiation. This is a perfectly consistent effective theory. The resulting BMS equation (not transversely ordered and multipole expanded according to the strict hard-soft factorization), looks very similar to the first BMS equation we derived, however with a subtly different boundary condition due to where the evolution is stopped and momentum regions used. We then proceed to find large phase-space logarithms in the boundary condition, which we argue would be resummed by the first formulation of the BMS equation in a CAESAR-like manner or by extending the types of momentum regions used in the SCET approach. 

Throughout, particular attention is paid to the ordering variable used in the BMS equation. In our view, this is immediately decided as soon as one accepts that the naive soft integrals determining the anomalous dimension must be split into on-shell and off-shell regions, where the off-shell region is governed by the Glauber/Coulomb gluon exchange. The splitting of the naive soft integral in this manner is natural in an effective theory with an explicit Coulomb potential mediating forward scattering effects. The split arises upon realizing the appropriate zero-bin subtractions, which manifests the ``Cheshire'' Glauber in Ref. \cite{Rothstein:2016bsq}. Then in order to not interfere with producing the correct imaginary part of the soft integral after this decomposition, and to maintain a trivial zero-bin subtraction, one must choose transverse-momentum ordering for the soft anomalous dimension.

We also work explicitly with renormalized quantities. In general, since we are ordering the emissions, this ordering parameter will regulate both IR and UV divergences. Within the SCET approach, after renormalizing the soft wilson lines, the ordering parameter is settled by the soft anomalous dimension. Ultimately, the IR divergences will cancel in the BMS equation, and the way we write the real insertions and virtual corrections will manifest how the cancellation can take place point-by-point in phase space \emph{when} the soft virtual corrections can be considered equivalent to their on-shell region.

\section{Reduced Density Matrix}
What is the reduced density matrix? Simply put, it is a way of organizing all the emissions above some resolution scale $\tau$, where this resolution scale is the scale at which the shower is terminated. We also assume that some hard process has taken place, with typical energy scale $Q$. All radiation below the $\tau$ scale according to some measure on momentum space is traced over, leading to the reduced density matrix. For instance, one could demand that no emission be counted as ``hard'' when it has a transverse momentum with respect to all other pairs of ``hard'' emissions below a scale $\tau$ (this is the now dominate Monte Carlo prescription). Alternatively, one could use an observable on the final state of the scattering like $N$-jettiness, a jet algorithm, or the energy-energy correlation functions to define the resolution scale and count hard states, as is done in Refs. \cite{Catani:2007vq,Stewart:2010tn,Alioli:2012fc,Larkoski:2013eya,Alioli:2013hqa,Gaunt:2015pea,Boughezal:2015dva,Larkoski:2015zka,Larkoski:2016zzc}. We then split all naive in and out states into hard and unresolved components. The hard state is composed of effective partons that themselves may be clusters of collinear radiation, having a dominate color, flavor, and spin. We then consider the unresolved states themselves to be states composing the effective hard parton, or soft radiation, classified according to our desired measure on momentum space:
\begin{align} 
|\sim\rangle&\rightarrow\text{hard}\otimes\text{unresolved}\\
|\alpha,in\rangle&=|A,in\rangle\otimes |r_\alpha,in\rangle\\
|m,out\rangle&=|M,out\rangle\otimes |r_m,out\rangle\\
|n,out\rangle&=|N,out\rangle\otimes |r_n,out\rangle
\end{align}
The density matrix formed by time evolving the initial state $|\alpha,in\rangle$ to $t=\infty$ is then:
\begin{align}
\hat{\rho}_{mn}&=\langle \alpha,in|m,out\rangle\langle n,out |\alpha,in\rangle
\end{align}
We now assume that the initial state is completely hard, for simplicity, so that then in the ''Fock space'' of all the emissions above the resolution scale, we can write hard reduced density matrix, organized by the emissions above the resolution scale:
\begin{align}
\hat{\rho}^H_{NM}(Q,\tau)&=\sum_{r, |r|<\tau}\langle \alpha,in|\Big(|M,out\rangle\otimes|r\rangle\langle r|\otimes\langle N,out |\Big)|\alpha,in\rangle \,.
\end{align}
The trace over the unresolved states is constrained by our resolution variable $\tau$. We have allowed the hard out-states to be off-diagonal, though the trace over the unresolved out-states leads to the eventual diagonalization of the directions of the hard partons. The diagonalization of the directions of the hard momenta follows from the fact the Kinoshita-Lee-Nauenberg theorem fails for the off diagonal elements, see Refs. \cite{PhysRevA.63.032102,Carney:2017jut}. However, though the hard states diagonalize in terms of the space-time path they take, they are not actually diagonal in all quantum numbers needed to specify a parton in the Fock space of emissions above the resolution scale. When we label the hard parton by its path taken in space-time, we use the result that a parton with large mass or large energy couples to softer radiation as an external current localized along a world-line, see for instance \cite{Collins:1981uk,Korchemsky:1985xj,Korchemsky:1991zp,Korchemsky:1992xv}. Tracing over all fluctuations about these world-lines of the soft radiation coupled to the path leads to the decoherence of superpositions of distinct world-lines, and the diagonalization of the space-time path. In the case of hard scattering, the world-lines point along the direction of the momentum of the hard parton exiting or entering the hard interaction region. However, the exact color state or spin state of the hard parton does not decohere due to the soft radiation, leading to the well-known fact that soft evolution induces a color rotation in the hard state on either side of a cut diagram. Physically, the resulting diagonal (in momenta) matrix elements are just proportional to the exclusive $N$-jet scattering cross-sections, where exclusive criteria is defined by the resolution scale used to carve out momentum space.

Generally, at leading-log and leading power, there will be some $N$ such that $\rho_{NN}(Q,\tau)$ is understood as the ``Born'' hard-process starting the shower, and the rest of the entries (with $N+1$ jets, $N+2$ jets, etc.) will have emissions populated by the shower starting from this seed. With the techniques of multi-jet-matching, see for instance Refs. \cite{Hoeche:2012yf,Gehrmann:2012yg,Frederix:2012ps,Platzer:2012bs,Alioli:2012fc,Lonnblad:2012ix}, one can include hard loop and radiative corrections (beyond the soft-and collinear corrections already produced by the parton-shower) to a potentially arbitrary number of the hard-scattering exclusive cross-sections.

We can also build the reduced density matrix within effective field theory using the so-called N-jet operators, see for instance Refs. \cite{Feige:2014wja,Larkoski:2014bxa,Beneke:2017ztn} (and references therein). These are formed from the basic building blocks:
\begin{align} \label{eq:soft_wilson_line}
\mathbf{S}_{n_i}&=\text{P exp}\Big(ig\int_0^{\infty}d\lambda n_i\cdot\mathbf{A}(\lambda n_i)\Big)\\
\label{eq:collinear_operators}X_{n_i}&\rightarrow\begin{cases}
B_{n_i\perp}^{\mu}=\frac{1}{g}[W_{n_i}^{\dagger}iD_{\perp}^{\mu}W_{n_i}]\\
\chi_{n_i}=W_{n_i}^{\dagger}\psi\\
\Phi_{n_i}=W_{n_i}^{\dagger}\phi^AT^AW_{n_i}
\end{cases}\\
W_n&=\text{P exp}\Big(ig\int_0^{\infty}d\lambda \bar{n}\cdot\mathbf{A}(\lambda \bar{n})\Big)
\end{align}
We should also consider past-point soft wilson lines. $X_{n_i}$ are field operators for either gluons, quarks, or scalars, and each collinear direction is parametrized by a light-cone direction it points along, $n_i$, and also the conjugate light-cone direction $\bar{n}_i$, satisfying $n_i\cdot \bar{n}_i\gg \frac{\tau}{Q}$. The $\perp$ denotes all directions transverse to the axis defined by $n_i$ and $\bar{n}_i$ in the rest frame of $n_i$ and $\bar{n}_i$. The $N$-jet operator is then:
\begin{align}\label{eq:N_jet_operator}
\mathcal{O}_N&=\mathbf{C}_N\Big(\{Q_in_i\}\Big)\otimes T\Big\{\prod_{i=1}^N\mathbf{S}_{n_i}\Big\}\otimes \prod_{i=1}^N \delta(Q_i-\bar{n}_i\cdot i\partial)X_{n_i}\,,
\end{align}
$\otimes$ is a convolution in the large momentum fractions $Q_i$, and $\mathbf{C}_N$ is the renormalized hard scattering amplitude, with IR divergences subtracted off and folded into the soft and collinear operators. So then the density matrix has the form:
\begin{align}
\rho_{NM}&=\mathcal{O}_N^{\dagger}\mathcal{O}_M+O\Big(\frac{\tau}{Q}\Big)\,.
\end{align}
The corrections are given by the expansion of the full theory scattering amplitudes about $N$ hard directions \cite{Larkoski:2014bxa}. It is now a simple matter to consider \emph{tracing} over specific collinear and/or soft sectors. The KLN theorem demands that at \emph{some} scale $\tau$, we must trace over all soft degrees of freedom.\footnote{Unless we can compute QCD scattering non-perturbatively, though soft photons will always require such a trace.} If we trace over the soft degrees of freedom, we will form the hard reduced density matrix, coupled to a soft function:
\begin{align}\label{eq:soft_trace}
&\rho_{NM}^H = \text{tr}_{s}\Big[\mathcal{O}_N^{\dagger}\mathcal{O}_M\Big]=C_N^{\dagger a_1a_2...a_N}\mathbf{S}_{n_1...n_N;n'_1...n'_M}^{a_1 b_1...a_N b_N;b'_1a_1'...b'_Ma'_M}(\tau)C_M^{a'_1...a_M'} \prod_{i=1}^NX_{n_i}^{b_i}\prod_{j=1}^MX_{n'_j}^{b'_j}\,,\\
&\mathbf{S}_{n_1...n_N;n'_1...n'_M}^{a_1 b_1...a_N b_N;b'_1a_1'...b'_Ma'_M}(\tau)=\sum_X \theta_\tau(X)\Big\langle 0\Big|T\{\prod_{i=1}^N\mathbf{S}_{n_i}^{a_ib_i}\}\Big|X\Big\rangle\Big\langle X\Big|\bar{T}\{\prod_{i=1}^M\mathbf{S}_{n'_i}^{\dagger b_i'a_i'}\}\Big|0\Big\rangle\,.
\end{align}
The trace over the soft states is constrained by our resolution variable $\tau$: we demand that the soft final state is consistent (inclusively) with the scale $\tau$. For instance, one could demand that the energy of each individual soft emission is below $\tau$:
\begin{align}
  \theta_{\tau}(X)=\prod_{p\in X}\theta(\tau-p^0)
\end{align} 
Within the SCET factorization literature, such soft functions are a generalization of the soft functions typically found in exclusive jet cross-sections, which are of the form:
\begin{align}
&\mathbf{S}_{n_1...n_N;n_1...n_N}^{a_1 b_1...a_N b_N;b_1a_1'...b_Ma'_M}(\tau)=\sum_X \theta_\tau(X)\Big\langle 0\Big|T\{\prod_{i=1}^N\mathbf{S}_{n_i}^{a_ib_i}\}\Big|X\Big\rangle\Big\langle X\Big|\bar{T}\{\prod_{i=1}^N\mathbf{S}_{n_i}^{\dagger b_ia_i'}\}\Big|0\Big\rangle\,.
\end{align}
Here, we have specifically diagonalized the number, color indices, and directions of the wilson lines on either side of the ``cut,'' but we have not diagonalized the color-indices that tie to the hard coefficient functions. This follows from our earlier argument that the tracing over soft degrees of freedom below a resolution scale will lead to a density matrix diagonal in the momentum eigenstates.

Armed with the reduced density matrix, we will want to calculate the expectation values of observables:
\begin{align}\label{eq:calculating_observables}
\langle O\rangle&=\text{tr}_H[\hat{\rho}^H(Q,\tau)O]=\sum_{N=0}^{\infty}\int_{\tau}^{Q} d\Phi_N \sum_{Q.N.} O_N\hat{\rho}^H_{NN}(Q,\tau)\,.
\end{align}
Where $d\Phi_N$ is the on-shell N-parton phase space, schematically integrated between the scales $Q$ (where the initiating hard process occurs) and the resolution scale $\tau$ that forms the lower limit to all resolved states.\footnote{This equivalent to integrating over the sliced phase-space, excluding regions that the measure produces a value less than $\tau$, exactly as in N-jettiness or $Q_T$ subtractions \cite{Catani:2007vq,Gaunt:2015pea,Boughezal:2015dva}.} $O_N$ is the value of the observable on the $N$-parton configuration. If we can take $\tau\rightarrow 0$ in Eq. \eqref{eq:calculating_observables}, then we have an infra-red and collinear safe observable. The sum denotes that we should sum (or average) over the quantum numbers of the hard emissions, like color, spin, flavor, potentially constrained by the measurement and initial conditions of the scattering. 

We wish to develop an integral equation that replaces the sum in Eq. \eqref{eq:calculating_observables} that will incorporate both initial and final state partons, and thus include the factorization violation effects examined in Refs. \cite{Catani:1985xt,Forshaw:2006fk,Forshaw:2008cq,Forshaw:2012bi,Catani:2011st,Angeles-Martinez:2015rna,Schwartz:2017nmr}. This evolution equation will be recognized as the Banfi-Marchesini-Smye equation \cite{Banfi:2002hw, Weigert:2003mm}, used to resum non-global soft correlations between distinct angular regions of phase space \cite{Dasgupta:2001sh,Dasgupta:2002bw}, suitably extended to account for initial state partons.

\section{Recursive Soft Gluon Insertions and Generating Functionals}\label{sec:coherent_branching}
We wish to formulate an object which will carry at the level of the amplitude all subsequent emissions off of a hard scattering. Within SCET, this is accomplished by just taking the appropriate matrix elements with soft and collinear states of the N-jet operator of Eq. \eqref{eq:N_jet_operator}. If instead of collinear operators and wilson lines as our fundamental building blocks, we would rather take the insertion of additional soft or collinear external states, generated by soft currents and splitting amplitudes, we can instead adopt a generating functional formalism. If $C_N$ is the renormalized hard scattering coefficient, we write:
\begin{align}
C_N\Big(p_1^{b_1\sigma_1},p_2^{b_2\sigma_2},...,p_N^{b_N\sigma_N}\Big)\rightarrow C_N\Big(p_1^{b_1\sigma_1},p_2^{b_2\sigma_2},...,p_N^{b_N\sigma_N}\Big)\mathcal{W}_{N}\Big(p_1^{b_1\sigma_1},p_2^{b_2\sigma_2},...,p_N^{b_N\sigma_N};U_i,U_f\Big) \text{ (no sum)}.
\end{align}
$\mathcal{W}_{N}$ carries all possible soft emissions and virtual corrections off of the hard state $\{p_1^{b_1\sigma_1},p_2^{b_2\sigma_2},...,p_N^{b_N\sigma_N}\}$, and will be given a precise definition below. Suffice it to say, it is a generating functional with functions $U_{i/f}$ that tie off all subsequent soft emissions. 

\subsection{Arbitrary Real Emissions}
Suppose we have a scattering amplitude $\mathcal{A}_{N}(q_1^{a_1\lambda_1},...,q_n^{a_n\lambda_n})$, where we have generated some soft final state $\{q_1^{a_1\lambda_1},...,q_n^{a_n\lambda_n}\}$ off of the hard state $\{p_1^{b_1\sigma_1},p_2^{b_2\sigma_2},...,p_N^{b_N\sigma_N}\}$. To connect with the algorithmic prescription of soft gluon insertions found in Ref. \cite{Bassetto:1982ma,Bassetto:1984ik,Catani:1984dp,Catani:1985xt,Fiorani:1988by,Mangano:1990by} (see as well \cite{Berends:1987me,Berends:1988zn,Catani:1999ss,Catani:2000pi,Li:2013lsa,Duhr:2013msa,Caron-Huot:2016tzz}), we need to develop an approximation to the full set of feynman diagrams encoded in $\mathcal{A}_{N}(q_1^{a_1\lambda_1},...,q_n^{a_n\lambda_n})$. At the leading log level, the heart of the approximation is just the tree-level soft factorization of amplitudes:\footnote{The lack of an $i0$ prescription is not a mistake, but is related to how we explicitly deal with the off-shell regions of integration in later sections. For those uncomfortable with this, just add a $+i0$ to the eikonal propagator.}
\begin{align}\label{eq:soft_gluon_insertion_rule}
\mathcal{A}_{N}(q_1^{a_1\lambda_1},...,q_n^{a_n\lambda_n})&=_{q_n^0\ll q_i^0}\mathcal{A}_{N}(q_1^{a_1\lambda_1},...,q_{n-1}^{a_{n-1}\lambda_{n-1}})\sum_{k\in\{p_1,...,p_N,q_1,...,q_{n-1}\}} g\mathbf{T}_k^{a_n}J_k^{(T)}(q_n)\cdot\epsilon_{q_n}(\lambda_{n})+...\,,\\
J_k^{(T)}(q)\cdot\epsilon_{q}(\lambda) &=\frac{k\cdot \epsilon_{q}(\lambda)}{k\cdot q}\,.
\end{align}
This is telling us that in the limit that the $n^{th}$ emission is softer than all the initial $N$ hard partons, as well as the $n-1$ subsequent emissions, the $n$ real emission amplitude factorizes into the amplitude for $n-1$ emissions and an eikonal factor for the $n^{th}$ emission. We sum over attachments of the $n^{th}$ soft gluon to all directions (the $N$ hard directions and the other (less) soft emissions). $\mathbf{T}^{a_n}_k$ is the color matrix in the representation of the parton k. This soft gluon insertion rule can be realized as a functional equation. To see how this is done, we first decompose the soft amplitude into a color basis (possibly over-complete, not necessarily orthogonal), while making the polarizations explicit. The color basis $\mathbf{C}_N^{[n]}$ is a color tensor in the $N$ by $n \otimes N$ color-space, mapping the color space of the initial hard $N$ directions to the $n\otimes N$ color-space of the final state. Written with explicit indices, it carries 2$N$-indices in the representations of the hard directions, and $n$ indices in the representations of the final state radiation.\footnote{We adopt the bracket $[]$ and parenthesis $()$ notation to explicitly denote the number of real emissions and virtual correctionsrespectively in a given object, as introduced in Ref. \cite{Larkoski:2014bxa}.}
\begin{align}\label{eq:soft_amplitude_in_color_basis}
\mathcal{A}_{N}^{[n]}&=\mathcal{A}_{N}(q_1^{a_1\lambda_1},...,q_{n}^{a_{n}\lambda_{n}})=\sum_{\mathbf{C}_N^{[n]}}C^{a_1...a_n}_{i_1 j_1,....,i_N j_N} J_{N}^{\mathbf{C}_N^{[n]}}(q_1^{\mu_1},...,q_{n-1}^{\mu_{n}})\prod_{i=1}^{n}\epsilon_{q_i}^{\mu_i}(\lambda_i)
\end{align}
One can recognize the $J_{N}^{\mathbf{C}_N}$ as the kinematic part of the soft currents for N-eikonal directions. For instance, for a single soft emission off an initial dipole, we would have:
\begin{align}\label{eq:soft_amplitude_in_color_basis}
\mathcal{A}_{2}^{[1]}&=\mathcal{A}_{2}(q_1^{a_1\lambda_1})=\sum_{\mathbf{C}_2^{[1]}}C^{a_1}_{i_1 j_1,i_2 j_2} J_{2}^{\mathbf{C}_2^{[1]}}(q_1^{\mu_1})\epsilon_{q_1}^{\mu_1}(\lambda_1)\nonumber\\
&=  g\left([\mathbf{T}_1^{a_1}]_{i_1j_1} \delta_{i_2j_2} \frac{p_{1{\mu_1}}}{p_1\cdot q}+[\mathbf{T}_2^{a_1}]_{i_2j_2} \delta_{i_1j_1} \frac{p_{2{\mu_1}}}{p_2\cdot q}\right)\epsilon_{q_1}^{\mu_1}(\lambda_1)
\end{align}
where the subscript i on $\mathbf{T}_i$ distinguishes the (possibly distinct) color representations for the initial partons created in the two hard directions.

With this decomposition, one ties to each emission a formal function. The function (which we call U) depends on whether the parton is in the initial state or the final state, and maps a null direction in momentum space, polarization, and color index to a (complex)-number. That is:\footnote{These functions can be extended, of course, to carry any other quantum number necessary to specify the parton's associated charges. Furthermore, for fermions we should take these functions to be anti-commuting variables. Since in what follows, we will simply work to leading log order, we ignore any charge or flavor indices.}
\begin{align}
&U_{i/f}: \mathbb{R}^{+}\otimes S^{2}\otimes\{+,-\}\otimes R[G]\rightarrow \mathbb{C}\,,\\
&\{q,a,\lambda\}\rightarrow U^{a\lambda}_{i/f}(q)\,.
\end{align}
Where $\mathbb{R}^{+}\otimes S^{2}$ are the energy and direction of the parton, $\{+,-\}$ the set of helicities, and $R[G]$ is its color representation. More explicitly, $a, \lambda$ are to be associated with the color index and polarization of a parton line with momentum $q=q^0(1,\hat{q})$, $\hat{q}^2=1$, and $i/f$ labels whether the parton is in the initial or final state.\footnote{Since we have assumed some hard scattering to have taken place, the time variable is always defined with respect to the total momentum of the initial state feeding into the hard scattering. This then also defines any center of mass frame.} We can then define a color tensor weighted by the $n$-point soft-amplitude, living in the $N$-index color-space of the initial hard lines by:
{\small\begin{align}\label{eq:group_space_soft_functional}
&\mathcal{W}_{N}^{[n]}\Big(U_i,U_f\Big)=\Big(\prod_{\ell=1}^NU^{j_\ell\lambda_{\ell}}_{z_\ell}(p_\ell)\Big)\sum_{a_i,\sigma_i}\Big(\prod_{i=1}^n\int [d^dq_i]_+U^{a_i\,\sigma_i}_f(q_i)\Big)\mathcal{A}_{N}(q_1^{a_1\sigma_1},...,q_{n}^{a_{n}\sigma_{n}})\,,\\
&z_\ell= i\text{ or } f \text{ for each } \ell\,,\\
\label{eq:on-shell-parton-phase-space}&[d^dp]_+=\frac{d^dp}{(2\pi)^{d-1}}\theta(p^0)\delta(p^2)=\frac{1}{4\pi}\frac{dp^0}{(p^0)^{3-d}}\theta(p^0)\frac{d^{d-2}\Omega_{\hat{p}}}{(2\pi)^{d-2}}\,.
\end{align}}
All color indices, the polarization indices, directions, and the energy dependence of any soft emission associated with the final state has also been integrated/summed over, being tied up by contracting with the appropriate $U$. Note that this object carries a single free color-index for each of the hard directions, and is itself an invariant color-tensor in the direct product space formed by the representations of the hard lines. We will call this object the color-space amplitude, since it maps the momenta and spins of the instigating eikonal lines to the hard color-space of those lines:
\begin{align}
\mathcal{W}_{N}^{[n]}\Big(U_i,U_f\Big): \otimes_{i=1}^N\Big(\mathbb{R}^{+}\otimes S^{2}\otimes\{+,-\}\Big)\rightarrow \otimes_{i=1}^{N} R_i\,,
\end{align} 
where $R_i$ is the representation of the $i$-th line. The formal meaning of these quantities is that we have contracted the open final state color, polarization, and momentum quantum numbers of the soft amplitude into the function $U$, affecting a type of functional moment transform to the hard color-space. For example, the color space amplitude for a final state dipole can be written as 
\begin{align}
\label{DipoleW}
\mathcal{W}_2^{[0]} &= U_f^{i_1 \lambda_1}(p_1) U_f^{i_2 \lambda_2}(p_2) =  U_f^{j_1 \lambda_1}(p_1) U_f^{j_2 \lambda_2}(p_2) \delta_{i_1j_1} \delta_{i_2j_2} \nonumber\\
\mathcal{W}_2^{[1]} &= U_f^{j_1 \lambda_1}(p_1) U_f^{i_2 \lambda_2}(p_2) \int [d^dq_1]_+U^{a_1\,\sigma_1}_f(q_1)  g[\mathbf{T}_1^{a_1}]_{i_1j_1} \frac{p_{1{\mu_1}}}{p_1\cdot q+i0}\epsilon_{q_1}^{\mu_1}(\sigma_1)\nonumber\\
&+U_f^{i_1 \lambda_1}(p_1) U_f^{j_2 \lambda_2}(p_2) \int [d^dq_1]_+U^{a_1\,\sigma_1}_f(q_1)  g[\mathbf{T}_2^{a_1}]_{i_2j_2} \frac{p_{2{\mu_1}}}{p_2\cdot q+i0}\epsilon_{q_1}^{\mu_1}(\sigma_1)
\end{align}
It is clear that this object (for N=2 and arbitrary soft emissions) has two free color indices and hence is a rank 2 tensor in color space. The subscript on the $\mathbf{T}$ matrix indicates the representation for the corresponding hard parton.

Through functional differentiation, the two representations of the soft amplitude are equivalent to each other. We define the functional derivative with respect to $U$ as:
{\small\begin{align} 
\frac{\delta}{\delta U^{a\lambda_1}_{z_1}(p)}U^{b\lambda_2}_{z_2}(q)&=2(2\pi)^{d-1}(p^0)^{3-d}\delta(p^0-q^0)\delta^{(d-2)}(\hat{p}-\hat{q})\delta^{\lambda_1\lambda_2}\delta^{ab}\delta_{z_1z_2}\,,
\end{align}}
Then demanding these derivatives obey the Leibniz rule gives a definition for these derivatives for any polynomial functional of the $U$'s. Note that the Kronecker delta for the color indices should also be interpreted as a Kronecker delta on the representation for those indices: $U$'s in distinct representations have a zero functional derivative. The delta functions for the momenta are defined with the appropriate jacobian for the fact we integrate over on-shell momenta in Eq. \eqref{eq:group_space_soft_functional}, that is:
\begin{align}
\frac{\delta}{\delta U^{a\lambda}_{i/f}(k)}\sum_{b,\sigma}\int[d^dq]_+U^{b\sigma}_{i/f}(q)f^{b\sigma}(q)&=f^{a\lambda}(k)\,.
\end{align}

\subsubsection{Color Space Generating Functional}
We are now in a position to write down the \emph{generating functional} for soft amplitudes. Since all the color-space amplitudes (with differing number of final state emissions n) live in the same hard color space (i.e., all have N free color indices), we can simply sum them to form:
\begin{align}\label{eq:color_space_gen_functional}
\mathcal{W}_{N}\Big(U_i,U_f\Big)&=\mathcal{W}_{N}^{[0]}\Big(U_i,U_f\Big)+\sum_{n_g,n_q=1}^{\infty}\frac{1}{n_g! n_q! n_{\bar q}!}\mathcal{W}_{N}^{[n_g,n_q,n_{\bar q}]}\Big(U_i,U_f\Big)\,.
\end{align}
This is the generating functional for all soft amplitudes off of $N$ initial hard lines, since any given soft momentum space amplitude is then constructed from the functional derivatives:
\begin{align}
\mathcal{A}_{N}(q_1^{a_1\lambda_1},...,q_{n}^{a_{n}\lambda_{n}})=\frac{1}{\prod_{\ell=1}^NU^{j_\ell\lambda_{\ell}}_{z_\ell}(p_\ell)}\prod_{i=1}^{n}\frac{\delta}{\delta U^{a_i\lambda_i}_f(q_i)}\mathcal{W}_{N}\Big(U_i,U_f\Big)\Bigg|_{U_{i/f}=0}\,.
\end{align}
Since the subsequent emissions are always softer than the initiating N hard lines, there is no interference between the U derivatives of hard and soft emissions. 
That is, going back to the color tensor decomposition of the soft amplitudes, we have the formal definition:
\begin{align}
\mathcal{W}_{N}\Big(U;\zeta\Big)&=\sum_{n=0}^{\infty}\frac{1}{n_g!n_q!n_{\bar{q}}!}\Big(\prod_{\ell=1}^NU^{j_\ell\lambda_{\ell}}_{z_\ell}(p_\ell)\Big)\sum_{a_i,\sigma_i}\Big(\prod_{i=1}^n\int [d^dq_i]_+U^{a_i\,\sigma_i}_f(q_i)\Big)\mathcal{A}_{N}(q_1^{a_1\sigma_1},...,q_{n}^{a_{n}\sigma_{n}})\,.
\end{align}
We note that each soft amplitude has a virtual loop expansion in $\alpha_s$;
\begin{align}\label{eq:loop_expansion_momentum_space_amplitudes}
\mathcal{A}_{N}(q_1^{a_1\lambda_1},...,q_{n}^{a_{n}\lambda_{n}})&=\mathcal{A}_{N}^{(T)}(q_1^{a_1\lambda_1},...,q_{n}^{a_{n}\lambda_{n}})+g^2\mathcal{A}_{N}^{(1)}(q_1^{a_1\lambda_1},...,q_{n}^{a_{n}\lambda_{n}})+g^4\mathcal{A}_{N}^{(2)}(q_1^{a_1\lambda_1},...,q_{n}^{a_{n}\lambda_{n}})+...
\end{align}
The generating functional we have constructed here thus far is a weighted sum of the complete QCD radiative corrections ( i.e., it includes the full phase space for arbitrary number of real and virtual corrections). In the subsequent analysis, we will primarily be interested in the strongly ordered limit of these corrections which will allow us to develop a recursive definition of $\mathcal{W}_N$.

\subsubsection{Digression on color index contractions}
We adopt a modified index notation for the color contractions of the color matrix $\mathbf{T}$, generalizing the prescription found in Refs. \cite{Catani:1996vz,Catani:1999ss}. If $C_N$ is a color tensor with $N$ free color indices:
\begin{align}
\mathbf{T}_i^{\ell}\circ C_{N}&=T_{a_i b_i}^{b_\ell}C_{a_1...a_{i-1}b_i\,a_{i+1}...a_{\ell-1}b_\ell a_{\ell+1}...a_N}\,,\\
C_{N}\circ\mathbf{T}_i^{\ell}&=C_{a_1...a_{i-1}b_i\,a_{i+1}...a_{\ell-1}b_\ell a_{\ell+1}...a_N}T_{b_i a_i}^{b_\ell}\,,\\
\mathbf{T}_i^{c} C_{N}&=T_{a_i b_i}^{c}C_{a_1...a_{i-1}b_i\,a_{i+1}...a_N}\,,\\
C_{N}\mathbf{T}_i^{c}&=C_{a_1...a_{i-1}b_i\,a_{i+1}...a_N}T_{b_i a_i}^{c}\,.
\end{align}
Thus the presence of the $\circ$ denotes that the index on the color matrix is contracted with the $\ell$-th color index. Without the $\circ$, this index is free. We will further adopt the notation:
\begin{align}
\mathbf{T}_i\circ\mathbf{T}_j\equiv T_{a_ib_i}^c T_{a_j b_j}^c\,.
\end{align}
For example, following Eq. \ref{DipoleW}, if we define 
\begin{align}
C_2 =  U_f^{i_1 \lambda_1}(p_1) U_f^{i_2 \lambda_2}(p_2)\,,
\end{align}
then we can write 
\begin{align}
\mathcal{W}_2^{[1]} &=   \int [d^dq_1]_+U^{a_1\,\sigma_1}_f(q_1) g\left(\frac{p_{1{\mu_1}}}{p_1\cdot q}\mathbf{T}_1^{a_1}+  \frac{p_{2{\mu_1}}}{p_2\cdot q}\mathbf{T}_2^{a_1}\right) C_2 \epsilon_{q_1}^{\mu_1}(\sigma_1)\,.
\end{align}
Equivalently, if we define 
\begin{align}
C_3 =  U_f^{i_1 \lambda_1}(p_1) U_f^{i_2 \lambda_2}(p_2)U^{a_1\,\sigma_1}_f(q_1) \,,
\end{align}
then we can write the same object as 
\begin{align}
\mathcal{W}_2^{[1]} &=   \int [d^dq_1]_+ g\left(\frac{p_{1{\mu_1}}}{p_1\cdot q}\mathbf{T}_1^{a_1}+  \frac{p_{2{\mu_1}}}{p_2\cdot q}\mathbf{T}_2^{a_1}\right)\circ C_3 \epsilon_{q_1}^{\mu_1}(\sigma_1)
\end{align}
So its clear the $\circ$ operation reduces the number of free color indices on $C_N$ by 1.

\subsection{The $LL$ Master Equation For Color Space Amplitudes}\label{sec:color_space_amplitudes}
The soft gluon insertion rule \eqref{eq:soft_gluon_insertion_rule} can be written as a functional derivative on the the color-space amplitude:
{\small\begin{align}\label{eq:soft_gluon_insertion_rule_II}
\mathcal{W}_{N}^{[n+1]}\Big(U_i,U_f\Big)&=\mathbf{J}^{(T)[1]}(U_i,U_f)\mathcal{W}_{N}^{[n]}\Big(U_i,U_f\Big)+...\,,\\
\label{eq:color_space_soft_insertion_op_1g}\mathbf{J}^{(T)[1]}(U_i,U_f)&=g\sum_R\int[d^dp]_+\int[d^dq]_+ \Big(J_p^{(T)}(q)\cdot\epsilon_q(\lambda) U^{a\lambda}_f(q)\Big)\,[\mathbf{T}_R^a]_{cd}\sum_{\sigma,z}U^{d\sigma}_z(p)\frac{\delta}{\delta U^{c\sigma}_{z}(p)}\,.
\end{align}}%
This formula is straightforward to interpret. We have dropped power corrections not corresponding to the strongly-ordered limit of the additional soft emission. The ``$\cdot$'' denotes a Lorentz index contraction. We sum over all the possible color representations $R$ of particles in the theory. The functional derivative acts on each $U$ in $\mathcal{W}_{N}^{[n]}\Big(U_i,U_f\Big)$ one-by-one, setting the color matrix $\mathbf{T}_R$ to the correct representation via the representation index $R$ (remembering the convention on Kronecker delta's of color indices are to be interpreted as Kronecker delta's on the representation also), setting the eikonal direction of the soft current $J_p^{(T)}$ to one of the final state or initial eikonal line directions, and ensuring the contraction with the appropriate color indices. Then the differentiated $U$ is replaced with itself again. Then another $U$ is added, corresponding to the additional soft gluon, formally contracting with its quantum numbers. Again, both $\mathcal{W}_{N}^{[n]}\Big(U_i,U_f\Big)$ and $\mathcal{W}_{N}^{[n+1]}\Big(U_i,U_f\Big)$ are color tensors in the \emph{same} color space. As an application, we can work out the first two iterations of the soft gluon insertion operator acting on a hard dipole $\mathcal{W}_{2}^{[0]}$. We will take all hard directions to be in the final state, so we may drop the initial/final label on $U$, giving:
{\small\begin{align}
&\mathcal{W}_{2}^{[0]}\Big(U\Big)= U^{i_1 \sigma_1}(p_1)U^{i_2 \sigma_2}(p_2)\,,\\
\label{eq:soft_current_on_dipole}&\mathbf{J}^{(T)[1]}(U)\mathcal{W}_{2}^{[0]}\Big(U\Big)=\sum_\lambda\int [d^dq]_+\Big(J_{p_1}^{(T)\mu}(q)T^{a}_{i_1j_1}\delta_{i_2j_2}+J_{p_2}^{(T)\mu}(q)T^{a}_{i_2j_2}\delta_{i_1j_1}\Big) U^{j_1 \sigma_1}(p_1)U^{j_2\sigma_2}(p_2)U^{a\lambda}(q)\epsilon^{\mu}_{q}(\lambda)\,,\\
\label{eq:two_soft_current_on_dipole}&\mathbf{J}^{(T)[1]}(U)\mathbf{J}^{(T)[1]}(U)\mathcal{W}_{2}^{[0]}\Big(U\Big)\nonumber\\
&\qquad=\sum_{\lambda,\rho}\int [d^dq]_+[d^dk]_+\Big(J_{p_1}^{(T)\mu}(q)J_{p_1}^{(T)\nu}(k)T^{a}_{i_1s_1}T^{b}_{s_1j_1}\delta_{i_2j_2}+J_{p_2}^{(T)\mu}(q)J_{p_2}^{(T)\nu}(k)T^{a}_{i_2s_2}T^{b}_{s_2j_2}\delta_{i_1j_1}\nonumber\\
&\qquad\qquad+J_{p_1}^{(T)\mu}(q)J_{p_2}^{(T)\nu}(k)T^{a}_{i_1j_1}T^{b}_{i_2j_2}+J_{p_2}^{(T)\mu}(q)J_{p_1}^{(T)\nu}(k)T^{a}_{i_1j_1}T^{b}_{i_2j_2}\nonumber\\
&\qquad\qquad+J_{p_1}^{(T)\mu}(q)J_{q}^{(T)\nu}(k)T^{c}_{i_1j_1}\delta_{i_2j_2}if^{acb}+ J_{p_2}^{(T)\mu}(q)J_{q}^{(T)\nu}(k)T^{c}_{i_2j_2}\delta_{i_1j_1}if^{acb}\Big)\epsilon^{\mu}_q(\lambda)\epsilon^{\nu}_k(\rho)\nonumber\\
&\qquad\qquad\qquad U^{j_1 \sigma_1}(p_1)U^{j_2 \sigma_2}(p_2)U^{j_1}(\hat{p}_1)U^{a\lambda}(k)U^{b\rho}(k)
\end{align}}%
One can recognize the result for one and two emissions strongly ordered off of the hard dipole. This reproduces for instance the strongly ordered limit of Eqn. 101 of Ref. \cite{Catani:1999ss}.

We are now in a position to state the chief result of this section: in the strongly ordered limit, any soft emission off of the seed eikonal lines can be promoted to an hard seed eikonal line in its own right. That is, there is an explicit relationship between the $N$ leading-log color-space amplitude and the $N+1$ leading-log color-space amplitude. This relationship we will call the \emph{master equation} for color-space amplitudes. We start with the initial condition:
\begin{align}\label{eq:gen_fun_initial_conditions}
&\mathcal{W}_{N}^{[0]}\Big(U_i,U_f\Big)=U^{j_1\lambda_1}_{z_1}(p_1)U^{j_2\lambda_2}_{z_2}(p_2)...U^{j_N\lambda_N}_{z_N}(p_N)\,.
\end{align}
We suppress the color and polarization indices on the left hand side for a more compact notation: the subscript $N$ denotes the number of open color indices $\mathcal{W}_{N}$ carries. Using the soft gluon operator $\mathbf{J}^{(T)[1]}$ repeatedly, we can write the leading-log approximation to the full color-space functional:
\begin{align}\label{eq:gen_fun_LL_real}
&\mathcal{W}_{N}^{[LL]}\Big(U_i,U_f\Big)\Big|_{real}=\mathcal{W}_{N}^{[0]}\Big(U_i,U_f\Big)+\sum_{n=1}^{\infty}\frac{1}{n!}\mathcal{W}_{N}^{[n]_{LL}}\Big(U_i,U_f\Big)\,,\\
&\mathcal{W}_{N}^{[n+1]_{LL}}\Big(U_i,U_f\Big)=\Big[\mathbf{J}^{(T)[1]}(U_i,U_f),\mathcal{W}_{N}^{[n]_{LL}}\Big(U_i,U_f\Big)\Big]\,.
\end{align}
We write this as a commutation relation so that the derivatives acting further to the right do not contribute, \emph{not} because of color. The current operator is constructed to be a color singlet (it has no free indices), and therefore it's color matrices commute. The subscript on the brackets, $[\sim]_{LL}$ is to highlight we have dropped power corrections in the strongly ordered limit. Therefore, we can simply exponentiate the soft gluon insertion operator:
\begin{align}\label{eq:gen_fun_LL_real_Expo}
&\mathcal{W}_{N}^{[LL]}\Big(U_i,U_f\Big)\Big|_{real}=\Bigg[\text{Exp}\Big[\mathbf{J}^{(T)[1]}(U_i,U_f)\Big],\mathcal{W}_{N}^{[0]}\Big(U_i,U_f\Big)\Bigg]\,.
\end{align}
We term this the leading-log $N$-eikonal line color-space amplitude generating functional. We also have the following functional differential equation, that relates the $N+1$ and the $N$-eikonal line leading log generating functionals via the master equation:
\begin{align}
\mathbf{U_{i/f}\cdot\frac{\delta}{\delta U_{i/f}}}&=_{def.}\int[d^dp]_+\sum_{a,\sigma}U^{a\sigma}_z(p)\frac{\delta}{\delta U^{a\sigma}_{i/f}(p)}\,.\\
\label{eq:gen_fun_LL_real_recursion}
\mathbf{U_{f}\cdot\frac{\delta}{\delta U_{f}}}\mathcal{W}_{N}^{[LL]}\Big|_{real}&=\sum_{i=1}^{N}\int[d^dp_{N+1}]\sum_{\sigma_{N+1}} \Big(J_{p_i}^{(T)}(p_{N+1})\cdot\epsilon_{p_{N+1}}(\sigma_{N+1}) \Big)\mathbf{T}^{N+1}_i\circ\mathcal{W}_{N+1}^{[LL]}\Big|_{real}
\end{align}
We note that $\mathcal{W}_{N+1}^{[LL]}$ depends on both the color index, momentum, and polarization of the $N+1$ lines, and we are formally contracting these into the current terms in Eq. \eqref{eq:gen_fun_LL_real_recursion}. Proof of this relation proceeds inductively on the number $n$ of inserted final state soft current operators, and we give in bullet form. 
\begin{itemize}
\item First we note:
{\small\begin{align}\label{eq:gen_fun_LL_real_recursion_induction}
&\mathbf{U_{f}\cdot\frac{\delta}{\delta U_{f}}} \mathcal{W}_{N}^{[n]_{LL}}\Big|_{real}=n\mathcal{W}_{N}^{[n]_{LL}}\Big|_{real}
\end{align}}
\item We state the induction hypothesis: 
{\small\begin{align}\label{eq:gen_fun_LL_real_recursion_induction}
&\mathcal{W}_{N}^{[n]_{LL}}\Big|_{real}=\sum_{i=1}^{N}\int[d^dp_{N+1}] \sum_{\sigma_{N+1}} \Big(J_{p_i}^{(T)}(p_{N+1})\cdot\epsilon_{p_{N+1}}(\sigma_{N+1})\Big)\mathbf{T}^{N+1}_i\circ\mathcal{W}_{N+1}^{[n-1]_{LL}}\Big|_{real}
\end{align}}
Note that the $N+1$ hard color index is to be contracted with the upper adjoint index $a_{N+1}$ implicit in $\mathbf{T}^{N+1}_i$.
\item Eq. \eqref{eq:gen_fun_LL_real_recursion_induction} is true for $n=1$.
\item Consider $n+1$:
{\small\begin{align}
&\mathcal{W}_{N}^{[n+1]_{LL}}\Big|_{real}=\Big[\mathbf{J}^{(T)[1]},\mathcal{W}_{N}^{[n]_{LL}}\Big|_{real}\Big]\\
&\quad=\Bigg[\mathbf{J}^{(T)[1]},\,\,\sum_{i=1}^{N}\int[d^dp_{N+1}] \sum_{\sigma_{N+1}}\Big(J_{p_i}^{(T)}(p_{N+1})\cdot\epsilon_{p_{N+1}}(\sigma_{N+1}) \Big)\mathbf{T}^{N+1}_i\circ\mathcal{W}_{N+1}^{[n-1]_{LL}}\Big|_{real}\Bigg]\\
&\quad=\sum_{i=1}^{N}\int[d^dp_{N+1}]\sum_{\sigma_{N+1}} \Big(J_{p_i}^{(T)}(p_{N+1})\cdot\epsilon_{p_{N+1}}(\sigma_{N+1}) \Big)\mathbf{T}^{N+1}_i\circ\mathcal{W}_{N+1}^{[n]_{LL}}\Big|_{real}
\end{align}}
The last line is true since the soft current operator only has derivatives that act on the $U$ functions, so that it commutes through the other terms.
\item The proof ends by differentiating the sum in Eq. \eqref{eq:gen_fun_LL_real}, and then trading the upper index for the lower index using Eq. \eqref{eq:gen_fun_LL_real_recursion_induction}.
\end{itemize}

\subsection{Ordered Virtual Soft Insertion Operator}
We now give the form of the operator that implements adding a virtual correction, again to one loop order. This virtual correction acts on the eikonalized lines, and hence its operation on \Eq{eq:color_space_soft_insertion_op_1g} does not produce full the one loop corrected soft current operator, but only the most strongly ordered region of such a correction. We have (repeated indices are summed):
{\small\begin{align}
\label{eq:color_space_soft_insertion_op_virt_1g}\mathbf{V}^{(1)}(U;\mu)&=\frac{1}{4}\sum_{R_1,R_2}\int[d^dp_1]_+ \int[d^dp_2]_+ \gamma_{12}(\mu)\,[\mathbf{T}_{R_1}^a]_{cd}U^{d\lambda_1}_{z_1}(p_1)\frac{\delta}{\delta U^{c\lambda_1}_{z_1}(\hat{p_1})}[\mathbf{T}_{R_2}^a]_{ef}U^{f\lambda_2}_{z_2}(p_2)\frac{\delta}{\delta U^{e\lambda_2}_{z_2}(\hat{p_2})}\,,\\
\gamma_{12}(\mu)&=4\pi\alpha_s(\mu)\Bigg(-\Gnorm i\pi\frac{\theta(z_1=z_2)}{4\pi^2}\nonumber\\
&\qquad+\int[d^4q]_+\theta\Big(\omega_1-\frac{n_2\cdot q}{n_1\cdot n_2}\Big)\theta\Big(\omega_2-\frac{n_1\cdot q}{n_1\cdot n_2}\Big)W_{12}(q)\,\mu\delta\Big(\mu-O(p_1,p_2;q)\Big)\Bigg)\,,\label{eq:soft_anom_dim_1_loop}\\
W_{ij}(q)&=J_{p_i}^{(T)}(q)\cdot J_{p_j}^{(T)}(q)=\frac{p_i\cdot p_j}{p_i\cdot q\,q\cdot p_j}\,,\\
\theta(\text{\bf{true}})&=1\,,\\
\theta(\text{\bf{false}})&=0\,,\\
p_i&=\omega_i(1,\hat{n}_i)\,,\qquad\hat{n}_i^2=1\,.
\end{align}}
The delta function of $\mu$ freezes the transverse integral, which if unrestricted is both UV and IR divergent. Eq. \eqref{eq:soft_anom_dim_1_loop} gives the soft anomalous dimension presented in App. \ref{app:anom_dim}, and utilizes the rewriting of the logarithm appearing in the soft anomalous dimension given in Eq. \eqref{eq:soft_logarithm}. We rewrite the logarithm in the dipole part of the soft anomalous dimension as an integral over an on-shell parton weighted by the soft eikonal factor with explicit cut-offs in energy. This may seem like an excessively complicated way to write a single logarithm, but it highlights how the virtual correction cancels the infra-red divergences of the real emissions point-by-point in phase space when we write out the BMS equation in Eq. \eqref{eq:BMS_equation_V1}. The $\theta$-functions with $\omega_1$ and $\omega_2$ correspond to cutoffs in the energy of the virtual emission. 

One may wonder why in a soft virtual integral, we have included cutoffs set by the energy of the emissions forming the dipole. They arise due to the contribution from collinear sectors to the soft anomalous dimension,\footnote{Recall that the \emph{soft} anomalous dimension controls the infra-red divergences of the hard scattering amplitude. Confusingly, in SCET, one would be liable to call it the \emph{hard} anomalous dimension.} that set the upper limit in rapidity to the eikonal integral. In the modern literature on the soft anomalous dimension, these collinear contributions are often realized by tilting all wilson lines off the light-cone, and subtracting eikonal jet functions, and adding in the full collinear sector contribution, see for instance Refs. \cite{Aybat:2006wq,Almelid:2017qju}. Our particular virtual operator can of course be upgraded to reproduce the full collinear anomalous dimension, so long as in the corresponding real emission terms we include contributions from the full collinear splitting kernels, but for the time being, we leave this aside.

The $i\pi$ term arises from exchange of coulomb/Glauber gluons. Since the soft currents as we have defined them \emph{do not} have a $+i0$ prescription, coupled with the fact that we have used the on-shell propagator $\delta(q^2)$ (recall Eq. \eqref{eq:on-shell-parton-phase-space}), we must explicitly include this $i\pi$ term. We have only integrated over the on-shell region of the \emph{virtual} soft parton. We can then interpret the anomalous dimension in the language of effective field theory as follows: integrating over the on-shell region only is equivalent to the naive soft region with a zero-bin subtraction (Ref. \cite{Manohar:2006nz}) in the Glauber region, removing the overlap to the off-shell region of integration. Then we explicitly add in the Glauber contribution via the Glauber Lagrangian insertion between the lines, see Refs. \cite{Rothstein:2016bsq,Schwartz:2017nmr}. This contribution is the Glauber exchange between two active lines connected to the hard interaction.

 We note that we have not specified the ordering function of the virtual emissions. This ordering must be the transverse momentum between the eikonalized lines. One can appeal to the long history within the Monte Carlo literature in Refs. \cite{Gustafson:1987rq,Skands:2009tb,Platzer:2009jq,Hartgring:2013jma} or explicit calculations using a dressed soft gluon insertion techniques (including Glauber effects) \cite{Forshaw:2006fk,Forshaw:2008cq,Forshaw:2012bi,Angeles-Martinez:2015rna} to justify this. For us, the proof is given in App. \ref{app:soft_virt}, and amounts to taking seriously the decomposition of the soft virtual integral into an purely soft (on-shell) region and a Glauber region, the overlap removed by a zero-bin subtraction. If we wish to \emph{not} perform such a decomposition of the naive soft integral, then we should adopt a transverse ordering to the naive soft integrals. So we conclude the proper ordering function should be:
\begin{align}\label{eq:transverse_ordering}
O(p_1,p_2;q)&=\sqrt{2\frac{p_1\cdot q\,q\cdot p_2 }{p_1\cdot p_2 }}=\sqrt{2W^{-1}_{12}(q)}\,.
\end{align}
For example in the case of a virtual exchange between two hard partons, one in the $n$ direction and the other in the $\bar n$ direction, we see that:
\begin{align}
O(n,\nbar;q) = \sqrt{ n \cdot q \bar n \cdot q} =  q_{\perp}\,,
\end{align}
where we have used the on-shell condition for $q$. The virtual emissions are therefore ordered in transverse momentum in the frame where the legs of the eikonal lines are back-to-back. To actually dress the generating functional with the full virtual correction, we will want to act on it with the integral of the virtual soft current operator. We can straightforwardly compute the action of this operator on $\mathcal{W}_{N}^{[0]}$:
\begin{align}
\int_{\mu_i}^{\mu_f}\frac{d\mu'}{\mu'}\mathbf{V}^{(1)}(U;\mu')\mathcal{W}_{N}^{[0]}(U)&=-\frac{1}{2}\sum_{1\leq i<j\leq N}\int_{\mu_i}^{\mu_f}\frac{d\mu'}{\mu'}\gamma_{ij}(\mu')\mathbf{T}_i\circ\mathbf{T}_j\mathcal{W}_{N}^{[0]}(U)\,.
\end{align}
The $\delta$ function in $O$ then automatically enforces ordering in transverse momentum. We can now dress the $\mathcal{W}_{N}^{[0]}$ with an arbitrary number of ordered virtual corrections, 
{\small\begin{align}\label{eq:virtual_exponetiation}
\mathcal{W}_{N}^{(LL)[0]}&(U;\mu_f,\mu_i)\Big|_{virtual}=\nonumber\\
&\mathcal{W}_{N}^{[0]}(U)+\int_{\mu_i}^{\mu_f}\frac{d\mu'}{\mu'}\Bigg[\mathbf{V}^{(1)}(U;\mu'),\mathcal{W}_{N}^{[0]}(U)\Bigg]\nonumber\\
&\qquad+\int_{\mu_i}^{\mu_f}\frac{d\mu'}{\mu'}\int_{\mu_i}^{\mu'}\frac{d\mu''}{\mu''}\Bigg[\mathbf{V}^{(1)}(U;\mu''),\Bigg[\mathbf{V}^{(1)}(U;\mu'),\mathcal{W}_{N}^{[0]}(U)\Bigg]\Bigg]\nonumber\\
&\qquad+\int_{\mu_i}^{\mu_f}\frac{d\mu'}{\mu'}\int_{\mu_i}^{\mu'}\frac{d\mu''}{\mu''}\int_{\mu_i}^{\mu''}\frac{d\mu'''}{\mu'''}\Bigg[\mathbf{V}^{(1)}(U;\mu'''),\Bigg[\mathbf{V}^{(1)}(U;\mu''),\Bigg[\mathbf{V}^{(1)}(U;\mu')\mathcal{W}_{N}^{[0]}(U)\Bigg]\Bigg]\Bigg]\nonumber\\
&\qquad+...
\end{align}}
We note that the soft virtual current operator with the largest $\mu$ acts first, and then the second largest acts, then the third largest, etc. The commutators ensure that the derivatives act no further to the right. This can be traced to the ordering of the indices on the color matrices in Eq. \eqref{eq:color_space_soft_insertion_op_virt_1g}. This way the first current to act will have it's $\mathbf{T}$ matrices as the outer most in the chain of inserted color generators. This setup by construction satisfies the differential equation:
\begin{align}\label{eq:Anomalous_dim_gen_fun}
\mu\frac{d}{d\mu}\mathcal{W}_{N}^{(LL)[0]}(U;\mu,\mu_i)\Big|_{virtual}&=-\frac{1}{2}\sum_{1\leq i<j\leq N}\gamma_{ij}(\mu)\mathbf{T}_i\circ \mathbf{T}_j\mathcal{W}_{N}^{(LL)[0]}(U;\mu,\mu_i)
\end{align}
This is solved by the path-ordered exponentiation of the kernel:
\begin{align}
\mathcal{W}_{N}^{(LL)[0]}(U;\mu_f,\mu_i)\Big|_{virtual}&=\mathcal{U}_{N}(\mu_f,\mu_0)\mathcal{W}_{N}^{(LL)[0]}(U;\mu_0,\mu_i)\Big|_{virtual}\\
\label{eq:Anomalous_dim_gen_fun_solve}\mathcal{U}_{N}(\mu_f,\mu_0)&=\mathcal{P}\text{Exp}\Bigg(-\frac{1}{2}\sum_{1\leq i<j\leq N}\int_{\mu_0}^{\mu_f}\frac{d\mu'}{\mu'}\gamma_{ij}(\mu')\mathbf{T}_i\circ\mathbf{T}_j\Bigg)
\end{align}
We also define $\mathcal{U}$ directly in terms of the soft anomalous dimension in Eq. \eqref{eq:soft_evo_factor}. We now wish to find a leading logarithmic approximation to the generating functional of Eq. \eqref{eq:color_space_gen_functional} for both the virtual and the real corrections. To combine the real and the virtual corrections together, we start our recursive definition of the leading log generating functional using a seed fully dressed with the virtual corrections:
\begin{align}\label{eq:gen_fun_LL}
&\mathcal{W}_{N}^{LL}\Big(U;\mu,\mu_i\Big)=\mathcal{W}_{N}^{(LL)[0]}\Big(U;\mu,\mu_i\Big)+\sum_{n=1}^{\infty}\frac{1}{n!}\mathcal{W}_{N}^{(LL)[n]_{LL}}\Big(U;\mu,\mu_i\Big)\\
&\mathcal{W}_{N}^{(LL)[n+1]_{LL}}\Big(U;\mu,\mu_i\Big)=\Big[\mathbf{J}^{(T)[1]}(U),\,\,\mathcal{W}_{N}^{(LL)[n]_{LL}}\Big(U;\mu,\mu_i\Big)\Big]
\end{align}
This is a seemingly trivial statement, since it looks like we merely drop the $[\,]'s$ about the $LL$. What will be important is that we will have the following modified relation between the $N$ and the $N+1$ leading log color-space amplitudes:
{\small\begin{align}\label{eq:gen_fun_LL_real_recursion_with_virt}
\mathbf{U_{f}\cdot\frac{\delta}{\delta U_{f}}} &\mathcal{W}_{N}^{LL}\Big(U;\mu,\mu_i\Big)=\nonumber\\
&\sum_{i=1}^{N}\int[d^dp_{N+1}]\sum_{\sigma_{N+1}} \Big(J_{p_i}^{(T)}(p_{N+1})\cdot\epsilon_{p_{N+1}}(\sigma_{N+1})\Big)\mathcal{U}_{N}(\mu,\mu_i)\mathbf{T}^{N+1}_i\circ\mathcal{U}_{N+1}^{-1}(\mu,\mu_i)\mathcal{W}_{N+1}^{LL}\Big(U;\mu,\mu_i\Big)
\end{align}}
This follows directly from Eq. \ref{eq:gen_fun_LL_real_recursion} by writing $\mathcal{W}_{N}^{LL}|_{\text{real}} = \mathcal{U}_{N}^{-1}(\mu,\mu_i)\mathcal{W}_{N}^{LL}\Big(U;\zeta;\mu,\mu_i\Big)$. Also the set of virtual diagrams that reproduce the leading log result are completely captured by dressing initial hard lines, i.e., at this order we are not sensitive to virtual corrections imposed on subsequent soft emissions off the initial hard lines.

\subsection{Ordered Real Soft Insertion Operator}
We can recursively dress the hard amplitude with both virtual corrections and real emissions using the virtual operator of Eq. \eqref{eq:color_space_soft_insertion_op_virt_1g}, and the current operator of \eqref{eq:soft_gluon_insertion_rule_II}. However, we have not yet implemented any ordering to the real current operator. Formally the soft real emissions could be at any scale, though we can imagine that the functions $U$ we have introduced have support at scales below the hard interaction. The conundrum we face is then how to order the real emissions explicitly. At the level of the amplitude, we only have the hard scales of the underlying process as the reference. That is, we could attempt to impose an ordering by dressing the current using:
\begin{align}
J_k^{(T)}(q)\cdot\epsilon_{q}(\lambda) \rightarrow \theta\Bigg(\mu-\frac{P^{ref}\cdot q}{\sqrt{s}} \Bigg) J_k^{(T)}(q)\cdot\epsilon_{q}(\lambda) \,.
\end{align}
$P^{ref}$ then must be some momentum associated with the hard process, for instance, we could take $P^{ref}=k$, the momentum of the hard leg, or $P^{ref}$ to be the total momentum of the initial state. The former would correspond to \emph{virtuality ordering} of the real emissions, the second would be an \emph{energy ordering}, which is known to be a problematic prescription, see Ref. \cite{Dokshitzer:2008ia}. However, since at the level of the amplitude, we do not yet know which leg will absorb the real emission when we square the amplitude, energy/virtuality ordering seems to be the only prescription which we could consistently implement if we insist on trying to write an evolution equation for the amplitude.

How might we order the real emissions in Eq. \eqref{eq:soft_gluon_insertion_rule_II}? Briefly, defining:
{\small\begin{align}
\mathbf{J}^{(T)[1]}(U_i,U_f;\mu)&=g\sum_R\int[d^dp]_+\int[d^dq]_+ \Big(J_p^{(T)}(q)\cdot\epsilon_q(\lambda) U^{a\lambda}_f(q)\Big)\,[\mathbf{T}_R^a]_{cd}\sum_{\sigma,z}U^{d\sigma}_z(p)\frac{\delta}{\delta U^{c\sigma}_{z}(p)}\mu\delta(\mu-n_p\cdot q).
\end{align}}
Where $n_p$ is the null vector in the direction of $p$, rescaled so $n^0_p=1$. We have then for the ordered real emissions:
\begin{align}
&\mathcal{W}_{N}^{[LL]}\Big(U_i,U_f;\mu,\lambda\Big)\Big|_{real}=\Bigg[P\text{exp}\Big[\int_{\lambda}^{\mu}\frac{d\mu'}{\mu'}\mathbf{J}^{(T)[1]}(U_i,U_f;\mu')\Big],\mathcal{W}_{N}^{[0]}\Big(U_i,U_f\Big)\Bigg]\,.
\end{align}

Ultimately, the ordered color-space amplitude will be of little use for us when calculating cross-sections, since we demand ordering with transverse momentum. We have developed it merely as an exercises in the generating functional technology. For a real emission, the transverse momentum is defined relative to the two eikonal lines, each on either side of the cut diagram. That is, while we can easily implement transverse ordering on the virtual corrections at the level of the amplitude, we must go to the level of the amplitude squared in order to efficiently define transverse ordering for the real emissions.

\subsection{The Amplitude Squared}
We now simply write down the correct evolution equation for the amplitude squared. We introduce a real emission operator, exactly analogous to Eq. \eqref{eq:color_space_soft_insertion_op_virt_1g}:
{\small\begin{align}
\label{eq:real_ordering}\mathbf{H}^{[1]}&(U,U^{\dagger};\mu)=\nonumber\\
&\frac{1}{2}\sum_{R_1,R_2}\int[d^dp_1]_+ \int[d^dp_2]_+ \mathbf{W}_{12}^{ab}(\mu;U,U^{\dagger})\,[\mathbf{T}_{R_1}^a]_{cd}U^{d\lambda_1}_{z_1}(p_1)\frac{\delta}{\delta U^{c\lambda_1}_{z_1}(\hat{p_1})}[\mathbf{T}_{R_2}^b]_{fe}U^{\dagger f\lambda_2}_{z_2}(p_2)\frac{\delta}{\delta U^{\dagger e\lambda_2}_{z_2}(p_2)}\,,\\
\mathbf{W}_{12}^{ab}&(\mu;U,U^{\dagger})=\sum_{\sigma,\lambda}\frac{4\pi\alpha_s(\mu)}{V}\int[d^dq]_+ W_{12}(q;\sigma,\lambda)\mu\delta\Big(\mu-O(p_1,p_2;q)\Big)U^{ a\sigma}_{f}(q)U^{\dagger b\lambda}_{f}(q)\,,\\
W_{ij}&(q;\sigma,\lambda)=J_{p_i}^{(T)}(q)\cdot \epsilon_q(\sigma)\epsilon^*_q(\lambda) \cdot J_{p_j}^{(T)}(q)=\frac{p_i\cdot \epsilon_q(\sigma)\epsilon^{*}_q(\lambda) \cdot  p_j}{p_i\cdot q\,q\cdot p_j}\,.
\end{align}}
$V$ is the normalizing factor. The definition and need for this factor will be explained in the next section when we calculate expectation values of observables using these generating functionals.

The action of $\mathbf{H}^{[1]}(U,U^{\dagger};\mu)$ is easy to calculate:
{\small\begin{align}
&\mathbf{H}^{[1]}(U,U^{\dagger};\mu)\mathcal{W}_{N}^{[0](0)}(U)\mathcal{W}_{N}^{\dagger[0](0)}(U)=\nonumber\\
&\qquad\sum_{1\leq i<j\leq N,\sigma}\frac{1}{V}\int[d^dp_{N+1}]_+ W_{ij}(p_{N+1})\mu\delta\Big(\mu-O(p_i,p_j;p_{N+1})\Big) \mathbf{T}_i^{N+1}\circ\mathcal{W}_{N+1}^{[0](0)}(U)\mathcal{W}_{N+1}^{\dagger[0](0)}(U)\circ\mathbf{T}_{j}^{N+1}\,.
\end{align}}
Here the color matrix $\mathbf{T}_i$ is inserted via $U$-derivatives, and the $\mathbf{T}_j$ via the $U^{\dagger}$ derivatives. And further, an additional $U$ and $U^\dagger$ functions are inserted, associated with the  additional final state emission. Since we are working in the strongly ordered limit, this emission now acts as a hard parton for all subsequent emissions which allows us to write the result in terms of $\mathcal{W}_{N+1}^{[0](0)}(U)\mathcal{W}_{N+1}^{\dagger[0](0)}(U)$. The real emission insertion operator $\mathbf{H}^{[1]}(U,U^{\dagger};\mu)$, like the virtual one is a color singlet and its action on $\mathcal{W}_{N}^{[0](0)}(U)\mathcal{W}_{N}^{\dagger[0](0)}(U)$ preserves the color space. 

We can now write down how to dress a hard amplitude squared with a arbitrary number of real and virtual emissions:
{\small\begin{align}
\label{eq:almost_BMS}\mathcal{W}_{N}^{[0](0)}(U)\mathcal{W}_{N}^{\dagger[0](0)}(U)\rightarrow \mathcal{P}\text{exp}\Bigg(\int_{\mu_S}^{\mu_H}\frac{d\mu}{\mu}\Big\{\mathbf{H}^{[1]}(U,U^{\dagger};\mu)+\mathbf{V}^{\dagger(1)}(U;\mu)+\mathbf{V}^{(1)}(U;\mu)\Big\}\Bigg)\mathcal{W}_{N}^{[0](0)}(U)\mathcal{W}_{N}^{\dagger[0](0)}(U)
\end{align}}
The operator is implemented in a $\mu$ ordered form in the same way as Eq. \ref{eq:virtual_exponetiation}. This implementation ensures the strongly ordered limit for both real and virtual corrections, i.e., the virtual corrections applied to a given real emission are automatically ordered with respect to the transverse momentum of $that$ emission. We only consider this subset of corrections since those are the ones which will recover the leading log result to all orders.

\subsection{Probabilities within the Generating Functional Approach}
To get to a probability, we need to square the amplitudes, and sum over all scatterings that contribute to the considered cross-section. This we do now, the following averaging rules for the functions $U$ and $U^{\dagger}$:
\begin{align}\label{eq:U_averaging}
\Big\langle U^{ a\lambda}_{x}(p) U^{\dagger b\sigma}_{y}(q)\Big\rangle&=2(2\pi)^{d-1}(p^0)^{3-d}\delta(p^0-q^0)\delta^{(d-2)}(\hat{p}-\hat{q})\delta^{\lambda\sigma}\delta^{ab}\delta_{xy}\,,\\
0&=\Big\langle U^{\dagger a\lambda}_x(p) U^{\dagger b\sigma}_{y}(q)\Big\rangle=\Big\langle U^{a\lambda}_x(p) U^{b\sigma}_{y}(q)\Big\rangle\,,\\
0&=\langle U^{\dagger a\lambda}_x(p)\rangle=\langle  U^{b\sigma}_{y}(q)\rangle\,.
\end{align}
We handle products of $U$'s Wick's Theorem.\footnote{This is because the $U$ averaging can be given a formal path integral definition, see Ref. \cite{ZinnJustin:1989mi}, which can be implemented numerically via an equivalent langevin simulation \cite{Hatta:2013iba,Hagiwara:2015bia}.} We can also introduce an averaging weighted by an observable:
{\small\begin{align}\label{eq:ave_with_obs}
\Big\langle U^{ a\lambda}_{x}(p) U^{\dagger b\sigma}_{y}(q)\Big\rangle_{\mathcal O}&=2(2\pi)^{d-1}(p^0)^{3-d}\delta(p^0-q^0)\delta^{(d-2)}(\hat{p}-\hat{q})\delta^{\lambda\sigma}\delta^{ab}\delta_{xy}\delta^{\lambda\sigma}O^{\lambda}_x(p)\,.
\end{align}}
We allow the observable to depend on whether the parton is in the initial state. Note that the $O$ can depend on all the parton momenta in this averaging rule. Nothing prevents us from including an arbitrary number of $U's$ in such an averaging rule, for instance, we might demand:
{\small\begin{align}\label{eq:ave_with_obs_II}
&\Bigg\langle\Big(\prod_{i=1}^{n} U^{ a_i \lambda_i}_{x_i}(p_i)\Big)\Big(\prod_{j=1}^{n} U^{\dagger b_j\sigma_j}_{y_j}(q_j)\Big) \Bigg\rangle_{O}= O\Big(\{p_i,\lambda_i,x_i\}_{i=1}^n\Big)\Bigg\langle\Big(\prod_{i=1}^{n} U^{ a_i \lambda_i}_{x_i}(p_i)\Big)\Big(\prod_{j=1}^{n} U^{\dagger b_j\sigma_j}_{y_j}(q_j)\Big) \Bigg\rangle\,.
\end{align}}
Where we average on the left hand side with no observable constraint. Which averaging rule we implement depends on the observable in question.

Now if we want to compute the soft contribution to a cross-section, we simply must calculate:
\begin{align}\label{eq:soft_trace_gen_func}
\mathbf{G}_N(O)&=\mathcal{N}\frac{\Big\langle \mathcal{W}_{N}(U_i,U_f)\mathcal{W}_{N}^{\dagger}(U_i,U_f)\Big\rangle_O}{\text{tr}\Big\langle \mathcal{W}_{N}(U_i,U_f)\mathcal{W}_{N}^{\dagger}(U_i,U_f)\Big\rangle}
\end{align}
The ratio is to cancel out the $\delta$-functions on the initial eikonal lines, and $\mathcal{N}$ is a normalization factor dictated by the normalization of the hard coefficients. Often we will suppress the denominator and the normalization, both being understood.

To illustrate the averaging rules, we work out the one loop real emission eikonal contribution to \Eq{eq:soft_trace_gen_func} from a color dipole, making the assumption that the measurement is independent of polarization, and all $U's$ are in the final state:
{\small\begin{align}
&\Big\langle \mathcal{W}_{2}^{[1]}(U)\mathcal{W}_{2}^{[1]\dagger}(U)\Big\rangle_O\nonumber\\
&\qquad= -4 \pi \alpha_s T^a_{i_1 j_1}T^a_{i_2j_2} \int_{\mu_i}^{\mu_f} \frac{d\mu}{\mu} \int [d^dq]_+ \frac{p_1 \cdot p_2}{p_1 \cdot q q\cdot p_2}\mu\delta\Big(\mu -\sqrt{2W^{-1}_{12}(q)}\Big)O(p_1,p_2,q)\frac{\langle U^{a\dagger}(q) U^a(q) \rangle}{V}
\end{align}}
The cross connection between the U's for the hard lines and soft emission is prevented by the fact that the polarization tensor for a massless particle is transverse to its momentum (we have gone ahead and performed the polarization sum produced by $U,U^\dagger$ averaging, further, we must remember the polarization sum produces a minus sign). The normalizing factor V is defined so as to remove the $\delta(0)$ terms that arise from the contraction of U and $U^{\dagger}$ at the same momentum. At higher order emissions, the ordering in $\mu$ prevents the contraction of U and $U^{\dagger}$ at different emission momenta. 


\subsection{The LL BMS Equation}\label{sec:BMS_Gen_Func}
We are now in a position to write out the leading log BMS equation. We simply take the soft trace in \Eq{eq:soft_trace_gen_func}, and substitute the leading log generating functional for the full generating functional using Eq. \eqref{eq:almost_BMS}:
{\small\begin{align}\label{eq:leading_log_gen_functional}
\mathbf{G}_N\Big(O;\mu_H,\mu_S\Big)= \Bigg\langle {\mathcal P}\text{exp}\Bigg(\int_{\mu_S}^{\mu_H}\frac{d\mu}{\mu}\Big\{\mathbf{H}^{[1]}(U,U^{\dagger};\mu)+\mathbf{V}^{\dagger(1)}(U;\mu)+\mathbf{V}^{(1)}(U;\mu)\Big\}\Bigg)\mathcal{W}_{N}^{\dagger[0](0)}(U)\mathcal{W}_{N}^{[0](0)}(U) \Bigg\rangle_O\,.
\end{align}}
Formally we should consider the limit $\mu_S\rightarrow 0$. We implicitly assume then that this limit exists. This is equivalent to demanding that the observable $O$ is infra-red and collinear safe. Otherwise we must set $\mu_S=\mu_{had} > \Lambda_{QCD}$, and between the scales $\mu_{had}$ and $\Lambda_{QCD}$ we will need a non-perturbative model.  It is important to note that the real and the virtual corrections have their integration regions foliated along contours constant $\mu$, so that the IR divergences will match point by point in phase space. \emph{This can be seen as the justification for Eq. \eqref{eq:leading_log_gen_functional}: this equation populates below the scale $\mu_H$ both real and virtual corrections from the $N$-hard eikonal lines. The virtual corrections match the leading order IR divergences in the virtual correction to the $N$ hard-lines, and the real correction matches the leading order IR divergences of one of the real emissions for $N+1$ hard amplitude squared, when that emission is taken soft.} Indeed, it was this observation that forms the basis of parton-showers based around Catani-Seymour subtractions \cite{Catani:1996jh,Catani:1996vz} or anntennae subtractions to form coherent dipole showers (Refs. \cite{Nagy:2006kb,Giele:2007di,Nagy:2007ty,Schumann:2007mg,Dinsdale:2007mf}): the subtraction procedure that renders the amplitude squared finite also tells one how to populate the next emission. The lowest order in perturbation theory is given as:
\begin{align}\label{eq:Lowest_Order_Boundary_Cond}
\mathbf{G}_N(O)&=O\Big(\{p_i,\lambda_i\}_{i=1}^N\Big)+...
\end{align}
which is simply the constraint giving the contribution of the N initial hard partons to the measurement. The Banfi-Marchesini-Smye equation simply controls the evolution of $\mathbf{G}_N\Big(O;\mu_H\Big)$ with changing initial hard scale $\mu_H$:\footnote{We have gone ahead and already performed the polarization sum in the BMS equation, so we must remember the polarization sum produces a minus sign on the real emission term.}
{\small\begin{align}\label{eq:BMS_equation_V1}
\mu_H\frac{d}{d\mu_H}&\mathbf{G}_N\Big(O;\mu_H\Big)=\nonumber\\
&-\frac{1}{2}\sum_{i,j=1}^{N} \frac{\alpha_s(\mu_H)}{\pi}\lambda_{ij}\Bigg[i\pi\frac{\mathbf{T}_{i}\circ \mathbf{T}_{j}}{2},\mathbf{G}_N\Big(O;\mu_H\Big)\Bigg]\nonumber\\
&-\sum_{i,j=1}^{N}2\pi \alpha_s(\mu_H)\int[d^4p_{N+1}]_{+} W_{ij}(p_{N+1})\mu_H\delta\Big(\mu_H-\sqrt{W_{ij}^{-1}(p_{N+1})}\Big)\nonumber\\
&\qquad\qquad\times\Bigg\{\mathbf{T}_{i}^{N+1}\circ\mathbf{G}_{N+1}\Big(O;\mu_H\Big)\circ\mathbf{T}_{j}^{N+1}\nonumber\\
&\qquad\qquad\quad-\theta\Big(\omega_i-\frac{n_j\cdot p_{N+1}}{n_i\cdot n_j}\Big)\theta\Big(\omega_j-\frac{n_i\cdot p_{N+1}}{n_i\cdot n_j}\Big)\Big(\frac{\mathbf{T}_{i}\circ \mathbf{T}_{j}}{2}\mathbf{G}_N\Big(O;\mu_H\Big)+\mathbf{G}_N\Big(O;\mu_H\Big)\frac{\mathbf{T}_{i}\circ \mathbf{T}_{j}}{2}\Big)\Bigg\}\,,\\
\lambda_{ij}&=\begin{cases}
1 \text{ if $i$ and $j$ are both initial or both final state partons\,,}\\ 
0 \text{ otherwise }
\end{cases}
\end{align}}
This equation follows in the same way as Eq. \ref{eq:Anomalous_dim_gen_fun}. The $\mathbf{G}_N$ and the $\mathbf{G}_{N+1}$ have distinct implicit phase-space constraints, as can be seen when using the averaging rules of Eq. \eqref{eq:ave_with_obs_II}. We have in the virtual on-shell terms a constraint from the energies $\omega_i$ and $\omega_j$ on the $n_i$ or $n_j$ components. We have not included such a constraint in the real emission term. This is justifiable to NLL order, as will be illustrated in Sec. \ref{sec:CAESAR}, as long as the averaging rules enforce the conservation of momentum for all the real emissions: conservation of momentum on $N+1$ emissions will produce these cut-offs anyways (but this would not happen for the virtual terms). We can also include these cut-offs in the real emission term explicitly as well, being redundant with the conservation of momentum, thus making the phase-space between the on-shell real and virtual emissions identical in the BMS equation.\footnote{This would be akin to using local conservation of momentum as done in some parton showers, for example see Refs. \cite{Giele:2007di,Platzer:2009jq}. In a local conservation scheme, the conservation of momentum is applied to both the no-splitting probability (virtual terms) and the real emissions, but only using the momenta of the currently decaying color-connected dipole, that is, the momenta of the two eikonal lines involved in the decay. As we will see in Sec. \ref{sec:CAESAR}, the difference between local and global momentum conservation is beyond NLL accuracy for double-log sensitive observables, unless we can prove some recoil-free conditions to all orders Refs. \cite{Bauer:2008dt,Larkoski:2014uqa}.} 

\subsection{The Out-of-Gap/Dressed Gluon Expansions and Calculating a Cross-Section}
One way to determine the nontrivial effects of these phase space-constraints is to use the Out-of-Gap or Dressed Gluon expansions of Refs. \cite{Forshaw:2006fk,Forshaw:2008cq,Forshaw:2012bi,Larkoski:2015zka,Larkoski:2016zzc}. This type of dressed-soft-jet expansion is also at the heart of every Monte-Carlo generator which weights the next emission by the Sudakov no-splitting kernels. This formally exponentiates all the virtual corrections to a certain perturbative order in the Sudakov exponent for \emph{each} real emission. All Monte Carlos then terminate after a finite number of real emissions, a prescription that converges to the correct solution of the BMS equation in all regions of phase-space \cite{Larkoski:2016zzc}. A similar claim cannot be made for the fixed order expansion. As noted in the beginning of Sec. \ref{sec:coherent_branching}, by construction, the $\mathcal{W}$ carries all subsequent emissions from the hard interaction, and $\mathbf{G}_N$ appropriately squares and averages over these amplitude level emissions. Thus if $C_{N}$ is the hard amplitude corresponding to the Born-level configuration of the hard scattering, $d\Phi_N$ is the on-shell phase space for the $N$ hard partons, then the observed cross-section is:
\begin{align}\label{eq:Cross_section}
\frac{d\sigma}{dO}&=\int d\Phi_N\lim_{\mu_{S}\rightarrow 0}\text{tr}[C_{N}\mathbf{G}_N\Big(O;\mu_H,\mu_{S}\Big)C_{N}^{\dagger}]\,,\\
d\Phi_N&=\prod_{i=1}^{N}[d^dp_i]_+
\end{align}
In this section we will develop these resummed expansions for $\mathbf{G}_N$. This will be a reorganization of the expansion for $\mathbf{G}_N$ not in terms of $\alpha_s$ but in terms of the number of real resolved emissions (\emph{jets}). One will include global/Sudakov resummation, and would correspond to a strict expansion in the number emissions, with each emission dressed by all its virtual corrections. This would be closely related to both how Monte Carlo event generators and the CAESAR resummation framework proceed. The dressed-gluon expansion will factor out the global or Sudakov resummation, since the global resummation depends on the details of the measurement about collinear regions. The residual equation will describe the non-global correlations between distinct angular regions. This factoring is of course not unique, one may factor any arbitrary constant between the ``global'' and the ``non-global'' contributions. Which logarithms to factor as part of the global resummation can be decided to all orders by looking at the born configuration of the observable's initial hard function, and collinear contribution to the observable. 

\subsubsection{Resummed Expansion}
To see how this reorganization can be made, we define a new object $\mathbf{g}_N$ which is related to $\mathbf{G}_N $ by a dressing of virtual corrections 
\begin{align}
\mathbf{G}_N\Big(O;\mu_H\Big)&=\mathcal{U}_N\Big(\mu_H;\mu_S\Big)\mathbf{g}_N\Big(\mu_H,\mu_S\Big)\mathcal{ U}_N^{\dagger}\Big(\mu_H;\mu_S\Big)\\
\end{align}
We also introduce the shorthand:
\begin{align}
d\Phi_{k}^{ij}(\mu)&=2\pi\alpha_s(\mu)(1-\delta^{ij})\int[d^4p_{k}]_{+} W_{ij}(p_{k})\mu\delta\Big(\mu-\sqrt{2W_{ij}^{-1}(p_{k})}\Big)\,,\\
\end{align}
This explicitly factors out the global hard evolution, $\mu_S$ is an arbitrarily low scale. We then plug this into Eq. \eqref{eq:BMS_equation_V1} which then gives us an evolution equation for $\mathbf{g}_N$:
{\small\begin{align}
\mu_H\frac{d}{d\mu_H}\mathbf{g}_N\Big(O;\mu_H,\mu_S\Big)=&-\sum_{i,j=1}^{N}\int d\Phi_{N+1}^{ij}(\mu_H)\Bigg\{\mathcal{ U}_N^{-1}\Big(\mu_H;\mu_S\Big)\mathbf{T}_{i}^{N+1}\circ\mathbf{G}_{N+1}\Big(O;\mu_H\Big)\circ\mathbf{T}_{j}^{N+1}\mathcal{U}_N^{-1\dagger}\Big(\mu_H;\mu_S\Big)\Bigg\}
\end{align}}
The boundary condition on $\mathbf{G}_N$ is the physical fact that we turn off the evolution at some IR scale $\mu_S$, so that:
\begin{align}
\mathbf{G}_N( O, \mu_H = \mu_S) =  O\Big(\{p_i,\lambda_i\}_{i=1}^N\Big)
\end{align}
We can now integrate both sides and use the boundary conditions to give:
{\small\begin{align}
\label{eq:BMS_for_dressed_gluons}\mathbf{G}_N&\Big(O;\mu_H,\mu_{S}\Big)=O\Big(\{p_i,\lambda_i\}_{i=1}^N\Big)\mathcal{U}_N^{\dagger} (\mu_H;\mu_{S})\mathcal{U}_N(\mu_H;\mu_{S})\nonumber\\
&-\sum_{i,j=1}^{N}2\pi \int_{\mu_{S}}^{\mu_H} \frac{d\mu}{\mu}\int d\Phi_{N+1}^{ij}(\mu)\, \mathcal{U}_N^{\dagger}(\mu_H,\mu)\mathbf{T}^{N+1}_i\circ\mathbf{G}_{N+1}\Big(O;\mu,\mu_{S}\Big)\circ \mathbf{T}_j^{N+1}\mathcal{U}_N(\mu_H,\mu)\,.
\end{align}}
Iterations of this equation then produce equation 1.5 of Ref. \cite{Angeles-Martinez:2015rna}, once we substitute in Eq. 1.1 found therein, or can be seen as generating the sum in Eq. \eqref{eq:calculating_observables} for the reduced density matrix calculation of the cross-section. 

For the high scale $\mu_H$, we should choose a point where the logarithms of the renormalized scattering amplitude is minimized. Naively, when we iterate Eq. \eqref{eq:BMS_for_dressed_gluons} and insert the resulting expansion into the cross-section of \eqref{eq:Cross_section}, we might expect a cancellation of the soft resummation factors $\mathcal{U}$. However, because of the measurement constraints, the integral up to $\mu_H$ in the second term of Eq. \eqref{eq:BMS_for_dressed_gluons} will be effectively cutoff above a certain scale. Thus the soft resummation factors when we insert this into the cross-sections using Eqs. \eqref{eq:Cross_section} will only cancel below the scale $m_H$. Thus as long as we take $\mu_S\ll m_H,m_L$, we are insensitive to the actual choice of $\mu_S$: this scale will cancel out order-by-order in the soft jet expansion.

\subsubsection{Illustration of a Measurement Constraint}
As an illustration of the measurement constraints $O\Big(\{p_i,\lambda_i\}_{i=1}^N\Big)$, we can take the example of the cumulative hemisphere thrust distributions in $e^+e^-\rightarrow $hadrons. The hemispheres are defined by the regions in which the total transverse momentum of all radiation in each hemisphere parallel to the plane dividing the hemispheres is zero. We then find (one minus) the thrust in each hemisphere. In the dijet limit, we can introduce light-cone directions $n=(1,\hat{t})$ and $\bar{n}=(1,-\hat{t})$, where $\hat{t}$ is in the direction perpendicular to the plane dividing the hemispheres, the thrust axis. Thus $\hat{t}$ gives the direction where most of the energy is flowing, and we can write the measurement function in this limit as:
{\small\begin{align}
O\Big(&\{p_i,\lambda_i\}_{i=1}^N\Big)=\nonumber\\
&\theta\Big(m_H-\sum_{i=1}^N n\cdot p_{i}\theta(\bar{n}\cdot p_{i}-n\cdot p_{i})\Big)\theta\Big(m_L-\sum_{i=1}^N\bar{n}\cdot p_i \theta(n\cdot p_{i}-\bar{n}\cdot p_{i})\Big)\nonumber\\
&\times\delta^{(2)}\Big(\sum_{i=1}^{n+2} p_{i\perp}\theta(\bar{n}\cdot p_{i}-n\cdot p_{i})\Big)\delta^{(2)}\Big(\sum_{i=1}^{n+2} p_{i\perp}\theta(n\cdot p_{i}-\bar{n}\cdot p_{i})\Big)\delta\Big(Q-\sum_{i=1}^{n+2}n\cdot p_i\Big)\delta\Big(Q-\sum_{i=1}^{n+2}\bar{n}\cdot p_i\Big)\,.
\end{align}}
$Q$ is the hard center of mass momentum of the $e^+e^-$ collision. We have used a projector to insure that the sum of transverse momenta in one hemisphere is zero, and momentum conservation gives the total momenta in the other hemisphere as necessarily zero. We have not performed any multipole expansion of the momentum conservation or measurement constraints, as would be done in a strict effective field theory approach. This is due to the fact that we have not, in our general evolution equation, specified an exact power counting for the soft radiation. We have also included the condition that the total transverse momentum in each hemisphere is zero, and our light-cone coordinate system is specified by the thrust axis $\hat{t}$:
\begin{align}
n=(1,\hat{t}), \quad\bar{n}=(1,\hat{t}),\quad n\cdot q_{\perp}=\bar{n}\cdot q_{\perp}=0
\end{align}
By conservation of momentum, we must have for all $N$ at least two of the momenta in the set $\{p_i\}_{i=1}^N$ satisfy (say $p_1$ and $p_2$):
\begin{align}
p_1&\approx \frac{Q}{2} n\,,\\
p_2&\approx \frac{Q}{2}\bar{n}\,.
\end{align}
That is more generally, within $\{p_i\}_{i=1}^N$, there must be two subsets of momenta which are within a cone of size $\sqrt{m_H/Q}$ and $\sqrt{m_L/Q}$ of each other, one parallel $\hat{t}$ and the other to $-\hat{t}$. These sets are carrying the bulk of the momenta, see Ref. \cite{Bauer:2008dt}.

\subsubsection{Recovering CAESAR}\label{sec:CAESAR}
We show how for an observable defined on an initial (time-like, so we can ignore the Glauber phase and possible factorization violating effects) color-dipole we can recover the CAESAR resummation formula (Ref. \cite{Banfi:2004yd})\footnote{For an extension of the CAESAR paradigm to NNLL for global observables, see Ref. \cite{Banfi:2014sua}, and see also Ref. \cite{Hoeche:2017jsi} for direct comparison to various Monte Carlo schemes of resummation.} up to sub-leading logarithmic effects. We adopt the shorthand for the phase-space and measurement constraint:
\begin{align}\label{eq:ordered_phase_space_shorthand}
d\phi_{k}^{ij}(\mu)&=2\pi\alpha_s(\mu)(1-\delta^{ij})\int[d^4p_{k}]_{+} W_{ij}(p_{k})\mu\delta\Big(\mu-\sqrt{2W_{ij}^{-1}(p_{k})}\Big)\,,\\
O\Big(\{p_i,\lambda_i\}_{i=1}^N\Big)&=O_N\,.
\end{align}
The Kronecker delta in Eq. \eqref{eq:ordered_phase_space_shorthand} will help condense the notation below. We then iterate Eq. \eqref{eq:BMS_for_dressed_gluons}:
{\small\begin{align}\label{eq:caesar_expansion_start}
\mathbf{G}_2&\Big(O;\mu_H,\mu_{S}\Big)=\nonumber\\
&O_2\mathcal{U}_2^{\dagger} (\mu_H,\mu_{S})\mathcal{U}_2(\mu_H,\mu_{S})\nonumber\\
&-\int_{\mu_{S}}^{\mu_H} \frac{d\mu}{\mu} \int d\phi_{3}^{i_3j_3}(\mu)O_3\mathcal{U}_2^{\dagger}(\mu_H,\mu)\mathbf{T}_{i_3}^{3}\circ\mathcal{U}_3^{\dagger} (\mu,\mu_{S})\mathcal{U}_3(\mu,\mu_{S})\circ\mathbf{T}_{j_3}^3\mathcal{U}_2(\mu_H,\mu)\nonumber\\
&+(-1)^2\int_{\mu_{S}}^{\mu_H} \frac{d\mu}{\mu} \int d\phi_{3}^{i_3j_3}(\mu)\int_{\mu_{S}}^{\mu} \frac{d\mu'}{\mu'}\int d\phi_{4}^{i_4j_4x}(\mu')\nonumber\\
&\qquad\qquad\qquad\qquad\qquad
O_4\mathcal{U}_2^{\dagger}(\mu_H,\mu)\mathbf{T}_{i_3}^{3}\circ\mathcal{U}_3^{\dagger} (\mu,\mu')\mathbf{T}^4_{i_4}\circ\mathcal{U}_4^{\dagger} (\mu',\mu_{S})\mathcal{U}_4(\mu',\mu_{S})\circ\mathbf{T}^4_{j_4}\mathcal{U}_3(\mu,\mu')\circ\mathbf{T}_{j_3}^3\mathcal{U}_2(\mu_H,\mu)\nonumber\\
&+...
\end{align}}
The repeated indices $i_n,j_n$ are summed over, each running through the set $\{1,....,n-1\}$. The definition of the phase space insures terms $i_n=j_n$ are set to zero. We now expand the resummation factors $\mathcal{U}$ with emission $p_3$ collinear to emissions $1$ or $2$, following App. \ref{sec:collinear_limit_anom_dim}. This is justifiable, since the measurement will constrain the invariant mass of $p_3$ with respect to $p_1$ or $p_2$ to be much smaller than the invariant mass $p_1\cdot p_2$. Then:
\begin{align}
\mathcal{U}_N(\mu_f,\mu_i)&=Sp_{13}(\mu_f,\mu_i)\mathcal{U}_{N-1}(\mu_f,\mu_i)\,,
\end{align}
We further use the fact that the initiating dipole is a color singlet, so that the exponent in $\mathcal{U}_2$, can be written as 
\begin{align}
2\mathbf{T}_1 \circ \mathbf{T}_2 = -\mathbf{T}_1^2 -\mathbf{T}_2^2
\end{align}
when acting on the two initiating hard partons. The R.H.S of this equation is just the Casimirs for the representation of the two partons and hence $\mathcal{U}_2$ (and its conjugate) can be pulled out of the overall expression.  This allows us to considerably simplify the expansion and we can see the first two terms become (focusing on the $1\parallel 3$ case):
{\small\begin{align}\label{eq:caesar_expansion_start}
\mathbf{G}_2&\Big(O;\mu_H,\mu_{S}\Big)=\nonumber\\
&O_2\mathcal{U}_2^{\dagger} (\mu_H,\mu_{S})\mathcal{U}_2(\mu_H,\mu_{S})\nonumber\\
&-\mathcal{U}_2^{\dagger} (\mu_H,\mu_{S})\mathcal{U}_2(\mu_H,\mu_S)\int_{\mu_{S}}^{\mu_H} \frac{d\mu}{\mu} \int d\phi_{3}^{i_3j_3}(\mu)O_3Sp_{13}^2(\mu,\mu_S)\mathbf{T}_{i_3}\cdot\mathbf{T}_{j_3}\nonumber\\
&+(-1)^2\int_{\mu_{S}}^{\mu_H} \frac{d\mu}{\mu} \int d\phi_{3}^{i_3j_3}(\mu)Sp_{13}^2(\mu,\mu_S)\int_{\mu_{S}}^{\mu} \frac{d\mu'}{\mu'}\int d\phi_{4}^{i_3'j_3'}(\mu')\nonumber\\
&\qquad\qquad\qquad\qquad O_4\mathcal{U}_2^{\dagger}(\mu_H,\mu')\mathbf{T}_{i_3}^{A_1}\mathbf{T}^{3}_{i_3'}\circ\mathcal{U}_3^{\dagger} (\mu',\mu_{S})\mathcal{U}_3(\mu',\mu_{S})\circ\mathbf{T}^3_{j_3'}\mathbf{T}_{i_3}^{A_1}\mathcal{U}_2(\mu_H,\mu')\nonumber\\
&+...
\end{align}}

Note that $\mathcal{U}_3$ in the second term will now only depend $p_2,\,p_4$ and on a null vector approximating $p_1+p_3$. The emission $p_4$ will only radiate off of the coherent sum of emissions $p_1$ and $p_3$, so that we relabel the indices $i_4$ and $j_4$ to $j_3'$ and $j_3'$ to emphasize that in this limit $p_4$ only radiates off of the dipole formed from $p_1+p_3$ and $p_2$. This dipole is again a color singlet which allows us to utilize the same simplification as before. We now expand $p_1\parallel p_4$:
{\small\begin{align}\label{eq:caesar_expansion_start}
\mathbf{G}_2&\Big(O;\mu_H,\mu_{S}\Big)=O_2\mathcal{U}_2^{\dagger} (\mu_H,\mu_{S})\mathcal{U}_2(\mu_H,\mu_{S})\nonumber\\
&-\mathcal{U}_2^{\dagger} (\mu_H,\mu_{S})\mathcal{U}_2(\mu_H,\mu_S)\int_{\mu_{S}}^{\mu_H} \frac{d\mu}{\mu} \int d\phi_{3}^{i_3j_3}(\mu)O_3Sp_{13}^2(\mu,\mu_S)\mathbf{T}_{i_3}\cdot\mathbf{T}_{j_3}\nonumber\\
&+(-1)^2\mathcal{U}_2^{\dagger} (\mu_H,\mu_{S})\mathcal{U}_2(\mu_H,\mu_{S})\int_{\mu_{S}}^{\mu_H} \frac{d\mu}{\mu} \int d\phi_{3}^{i_3j_3}(\mu)Sp_{13}^2(\mu,\mu_S)\nonumber\\
&\qquad\qquad\qquad\qquad\qquad\qquad\qquad\int_{\mu_{S}}^{\mu} \frac{d\mu'}{\mu'}\int d\phi_{4}^{i_3'j_3'}(\mu') Sp_{14}^2(\mu',\mu_S)O_4\mathbf{T}_{i_3}^{A_1}\mathbf{T}^{A_2}_{i_3'}\mathbf{T}^{A_2}_{j_3'}\mathbf{T}_{i_3}^{A_1}\nonumber\\
&+...
\end{align}}
We then expand in the strongly ordered collinear limit, so that emission $p_{N+1}$ is collinear to $p_1$, after we have taken all the previous emissions in that limit. The momentum $p_3$ will be collinear to either $p_1$ or $p_2$, and that will decide the collinearity of all subsequent emissions.  We can also treat the initial momenta $p_1$ and $p_2$ to be back-to-back, so that expanding $p_3$ to be collinear to $p_1$ or $p_2$ will result in the same eikonal factor. We do not yet explicitly expand the measurement constraints (a point which is crucial for getting a finite result), only the resummation factors and eikonal factors. We then have the result when we also include the region where each emission is parallel to $p_2$:
{\small\begin{align}\label{eq:caesar_expansion_start}
&\mathbf{G}_2\Big(O;\mu_H,\mu_{S}\Big)=\nonumber\\
&\quad\mathcal{U}_2^{\dagger} (\mu_H,\mu_{S})\mathcal{U}_2(\mu_H,\mu_{S})\Bigg(O_{2}\nonumber\\
&\quad\quad+\sum_{n=1}^{\infty}(\frac{\mathbf{T}_1^2+\mathbf{T}_2^2}{2})^n\mathcal{P}\int_{\mu_{S}}^{\mu_H}\prod_{i=1}^{n} \frac{d\mu_i}{\mu_i} \int d\phi_{i+2}^{12}(\mu_i)\Big(O_{n+2}Sp_{1\,i+2}^2(\mu_i,\mu_S)\Bigg|_{\parallel p_1}+O_{n+2}Sp_{2\,i+2}^2(\mu_i,\mu_S)\Bigg|_{\parallel p_2}\Big)\Bigg)
\end{align}}%
where we have the path-ordered integration:
\begin{align}
\mathcal{P}\int_{\mu_{S}}^{\mu_H}\prod_{i=1}^{n} \frac{d\mu_i}{\mu_i} &=\int_{\mu_{S}}^{\mu_H}\frac{d\mu_1}{\mu_1}\int_{\mu_{S}}^{\mu_1}\frac{d\mu_2}{\mu_2} ...\int_{\mu_{S}}^{\mu_{n-1}}\frac{d\mu_n}{\mu_n}\,.
\end{align}
Eq. \eqref{eq:caesar_expansion_start} is essentially the CAESAR resummation formula. In order to achieve full next-to-leading log accuracy claimed for the CAESAR formalism, we must however use the running coupling in Eq. \eqref{eq:ordered_phase_space_shorthand} in the CMW scheme \cite{Catani:1990rr,Dokshitzer:1995ev}, and instead of using the eikonal factor $W_{12}(p)$, we should promote the eikonal factor to include the collinear limits as well, by using the appropriate antenna or Catani-Seymour functions for radiation off of a quark or gluon pair. If we take the example of the total cumulative thrust distribution, we then have for the measurement function:
{\small\begin{align}
O_{n+2}=\theta\Big(\tau&-\sum_{i=1}^{n+2} n\cdot p_{i}\theta(\bar{n}\cdot p_{i}-n\cdot p_{i})-\sum_{i=1}^{n+2}\bar{n}\cdot p_i \theta(n\cdot p_{i}-\bar{n}\cdot p_{i})\Big)\nonumber\\
&\times\delta^{(2)}\Big(\sum_{i=1}^{n+2} p_{i\perp}\theta(\bar{n}\cdot p_{i}-n\cdot p_{i})\Big)\delta^{(2)}\Big(\sum_{i=1}^{n+2} p_{i\perp}\theta(n\cdot p_{i}-\bar{n}\cdot p_{i})\Big)\delta\Big(Q-\sum_{i=1}^{n+2}n\cdot p_i\Big)\delta\Big(Q-\sum_{i=1}^{n+2}\bar{n}\cdot p_i\Big)\,.
\end{align}}
Given that we can take $p_1$ and $p_2$ to lie in separate hemispheres, when integrating over the transverse momentum and the longitudinal momentum fraction of $p_1$, we are really integrating over all possible orientations of the axis for the hemispheres. Each configuration will give an identical thrust distribution. We also use the symmetry between each hemisphere to remap the integration of each emission to be solely in the $p_1$ hemisphere. Thus we achieve:
{\small\begin{align}\label{eq:caesar_thrust}
\int d\bar{n}\cdot p_1\int d^2p_{1\perp}&\mathcal{P}\int_{\mu_{S}}^{\mu_H}\prod_{i=1}^{n} \frac{d\mu_i}{\mu_i} \int d\phi_{i+2}^{12}(\mu_i)O_{n+2}\Bigg|_{\parallel p_1}\nonumber\\
&=\Big(\frac{2}{\pi}\Big)^n\mathcal{P}\int_{\mu_{S}}^{\mu_H}\prod_{i=1}^{n} \frac{d\mu_i}{\mu_i}\alpha_s(\mu_i)\int_{\frac{\mu_i}{Q}}^1\frac{dz_i}{z_i} \theta\Big(\tau-\sum_{i=1}^{n}\frac{\mu_i^2}{Qz_i}\Big)\theta(1-\sum_{i=1}^{n}z_i)\
\end{align}}
here we have used the transverse ordering and on-shell condition to write:
\begin{align}
n\cdot p_{i+2}&=\frac{\mu_i^2}{\bar{n}\cdot p_{i+2}}\,,\qquad \bar{n}\cdot p_{i+2}=Qz_i \\
\int d\phi_{i}^{12}(\mu_i)&=\frac{1}{2\pi}\frac{\alpha_s(\mu_i)}{\mu_i}\int\frac{dz_i}{z_i}\,.
\end{align}
Formally, $p_1$ will contribute to the thrust with a term:
\begin{align}
n\cdot p_1=-\frac{|\sum_{i=3}^{n+2}\vec{p}_{i\perp}|^2}{Q(1-\sum_{i=3}^nz_i)}\,.
\end{align}
But within the region where $\mu_i\sim Q z_i \ll \mu_H$, this term is sub-dominate. Then using:\footnote{This approximation for integrating over the full conservation of the large momentum is what justifies not treating the phase space of the real and virtual emissions identically in Eq. \eqref{eq:BMS_equation_V1}.}
\begin{align}
\prod_{i=1}^{n}\int_{x_i}^\infty\frac{dz_i}{z_i}\theta\Big(A-\sum_{i=1}^{n}z_i\Big)&=\prod_{i=1}^{n}\int_{x_i}^A\frac{dz_i}{z_i}+O(N^2LL)\,,
\end{align}
where by $N^2LL$ we mean that we explicitly loose two logarithms. Then we have:
{\small\begin{align}
\text{Eq. \eqref{eq:caesar_thrust}}=\Big(\frac{2}{\pi}\Big)^n&\mathcal{P}\int_{\mu_{S}}^{\mu_H}\prod_{i=1}^{n} \frac{d\mu_i}{\mu_i}\alpha_s(\mu_i)\int_{\frac{\mu_i}{Q}}^{1}\frac{dz_i}{z_i} \theta\Big(\tau-\sum_{i=1}^{n}\frac{\mu_i^2}{Qz_i}\Big)+O(N^2LL)\,.
\end{align}}
To finish, we can take a derivative with respect to $\tau$, set $\mathbf{T}_1^2=\mathbf{T}_2^2=C$, and Laplace transform to achieve:
{\small\begin{align}
\int_0^{\infty}d\tau e^{-\theta\tau}\frac{d}{d\tau}\mathbf{G}_2\Big(O;\mu_H,\mu_{S}\Big)&=\mathcal{U}_2^{\dagger} (\mu_H,\mu_{S})\mathcal{U}_2(\mu_H,\mu_{S})\text{exp}\Bigg(C\frac{4}{\pi}\int_{\mu_{S}}^{\mu_H}\frac{d\mu}{\mu}\alpha_s(\mu)\int_{\frac{\mu}{Q}}^{1}\frac{dz}{z}e^{-\theta\frac{\mu^2}{Qz}}\Bigg)\,.
\end{align}}
We have for the resummation from the soft anomalous dimension:
{\small\begin{align}
\mathcal{U}_2^{\dagger} (\mu_H,\mu_{S})\mathcal{U}_2(\mu_H,\mu_{S})&=\text{exp}\Bigg(-C\frac{2}{\pi}\int_{\mu_{S}}^{\mu_H}\frac{d\mu}{\mu}\alpha_s(\mu)\int_{\frac{\mu^2}{Q^2}}^{1}\frac{dz}{z}\Bigg)\,.
\end{align}}
The upper and lower limits on the virtual momentum fraction integral as set by the momentum conservation constraint. We then use:
\begin{align}
\int_{\frac{\mu^2}{Q^2}}^{1}\frac{dz}{z}&=2\int_{\frac{\mu}{Q}}^{1}\frac{dz}{z}\,,
\end{align}
to get our final result:
{\small\begin{align}
\int_0^{\infty}d\tau e^{-\theta\tau}\frac{d}{d\tau}\mathbf{G}_2\Big(O;\mu_H,\mu_{S}\Big)&=\text{exp}\Bigg(C\frac{4}{\pi}\int_{0}^{\mu_H}\frac{d\mu}{\mu}\alpha_s(\mu)\int_{\frac{\mu}{Q}}^{1}\frac{dz}{z}\Big(e^{-\theta\frac{\mu^2}{Qz}}-1\Big)\Bigg)\,.
\end{align}}
We now use the approximation from Ref. \cite{Catani:1992ua}:
\begin{align}
e^{-\theta\,a}-1\approx -\theta\Big(a-(e^{\gamma_E}\theta)^{-1}\Big)
\end{align}
Then we split up the integration regions as:
{\small\begin{align}
\theta\Big(\frac{\mu^2}{Qz}-(e^{\gamma_E}\theta)^{-1}\Big)\theta(1-z)\theta\Big(z-\frac{\mu}{Q}\Big)&=\theta\Big(e^{\gamma_E}\theta\frac{\mu^2}{Q}-1\Big)\theta(1-z)\theta\Big(z-\frac{\mu}{Q}\Big)\nonumber\\
&\quad+\theta\Big(1-e^{\gamma_E}\theta\frac{\mu^2}{Q}\Big)\theta\Big(e^{\gamma_E}\theta\mu-1\Big)\theta\Big(e^{\gamma_E}\theta\frac{\mu^2}{Q}-z\Big)\theta\Big(z-\frac{\mu}{Q}\Big)
\end{align}}
So that:
{\small\begin{align}
\int_{0}^{\mu_H}\frac{d\mu}{\mu}\alpha_s(\mu)\int_{\frac{\mu}{Q}}^{1}\frac{dz}{z}\Big(e^{-\theta\frac{\mu^2}{Qz}}-1\Big)&=\int_{\mu_J(\theta)}^{\mu_H}\frac{d\mu}{\mu}\alpha_s(\mu)\text{ln}\frac{Q}{\mu}-\int_{\mu_S(\theta)}^{\mu_J(\theta)}\frac{d\mu}{\mu}\alpha_s(\mu)\text{ln}(\mu\theta\,e^{\gamma_E})+...\\
\mu_{J}(\theta)&=\sqrt{\frac{Q}{\theta\,e^{\gamma_E}}}\label{eq:emergent_collinear_scale}\\
\mu_{S}(\theta)&=\frac{1}{\theta\,e^{\gamma_E}}\label{eq:emergent_soft_scale}
\end{align}}
This is to be compared to the SCET result for the resummation of thrust. In laplace-space, the scale setting choices for the renormalization group evolution are precisely given by Eqs. \eqref{eq:emergent_collinear_scale} and \eqref{eq:emergent_soft_scale} see Refs. \cite{Schwartz:2007ib,Fleming:2007qr,Almeida:2014uva}, corresponding to the typical collinear opening angle and ultra-soft energy scale for the thrust distribution.

\section{Soft Amplitudes and Soft Functions}

In this section we will examine the construction of the BMS equation using the definition of soft functions from wilson lines\footnote{See references \cite{Becher:2015hka,Becher:2016mmh} for another SCET based approach.}. We will give the formal definition of a soft trace for additive observables on $N$-directions, corresponding to an insertion of a complete set of states normally used to calculate a soft function. The ingredients for this soft trace are writing down an operator definition of the generating functional for the soft emissions, valid to arbitrary order. Then using Lehmann-Symanzik-Zimmerman (LSZ) reduction formula to accomplish the insertion of the complete set of states, we write down the form of the measurement operator that generates a state consistent with the required final state constraints. This operator admits a recursive description of ordered emissions in energy, to any given logarithmic accuracy. Then following the logic of \cite{Weigert:2003mm}, but written in terms of objects familiar to an effective field theorist, we will give a leading logarithmic evolution equation, the BMS equation, for this soft trace, using this recursively defined insertion of ordered states. The method of proof is straightforward, and extends to higher orders, being limited only by the calculated perturbative accuracy of soft currents. This directly connects the BMS equation as an evolution equation satisfied by the standard effective field theory definitions of soft functions.

When building the BMS evolution equation, we will use a minimally consistent power counting, as done in Ref. \cite{Becher:2016mmh} in the example of hemisphere jet observables. This power counting is sufficient to resum all non-global logarithms, but does not suffice to resum all large phase-space logarithms, as we will illustrate. 

\subsection{Soft amplitudes and soft functions from Wilson lines} 
A basic building block for the analysis of soft physics is the vacuum matrix element of time-ordered wilson lines. To this end we introduce the soft amplitude functional:
\begin{align}\label{eq:Soft_Amplitude_Functional}
\mathbf{Y}_N[j]&=\langle 0|T\Big\{\mathbf{S}_{p_1}...\mathbf{S}_{p_N}e^{iS[A]+i j\cdot A}\Big\}|0\rangle\,,\\
j\cdot A&=_{df}\int d^dx j^{a}_{\mu}(x)A^{a\mu}(x)\,.
\end{align}
Here we have introduced a current only for the gluon field, but all parton flavors would have their distinct currents, which we suppress for conciseness. $S$ is simply the action for the gauge theory under consideration. This functional in principle encodes all the possible soft amplitudes. The $n$-point soft amplitudes are the amplitudes for the production of $n$-partons from the fixed number of hard scatters. These soft amplitudes are defined via the LSZ reduction procedure on the soft amplitude functional:
\begin{align}\label{eq:LSZ_reduction_Amplitude}
\mathcal{A}_{N}^{[n]}(q_1^{a_1\lambda_1},...,q_n^{a_n\lambda_n})&=\prod_{i=1}^n\Bigg(\lim_{q_i^2\rightarrow 0}\int d^dx_ie^{iq_i\cdot x_i}\epsilon_{q_i}^{\mu_i}(\lambda_i)\partial_{x_i}^2\frac{\delta}{i\delta j^{a_i\mu_i}(x_i)}\Bigg)\mathbf{Y}_N[j]\Bigg|_{j=0}\,.
\end{align}
Where $\epsilon_{q_i}^{\mu_i}(\lambda_i)$ is the polarization vector for the $i$-th soft parton, $q_i$ its momentum, $a_i$ its color index, and $\lambda_i$ the polarization quantum number ($\pm$ for helicity). At this stage, all the emissions are softer than the initiating N hard partons, but otherwise arbitrary. This amplitude also includes all order virtual corrections for each real emission which is obtained by successive insertions of the Lagrangian interaction in the exponent. To condense notation, we will denote the LSZ reduction operator as:
\begin{align}
L\Big(q_{1}^{a_1\lambda_1},...,q_{n}^{a_n\lambda_n};\frac{\delta}{\delta j}\Big)=\prod_{i=1}^n\Bigg(\lim_{q_i^2\rightarrow 0}\int d^dx_ie^{iq_i\cdot x_i}\epsilon_{q_i}^{\mu_i}(\lambda_i)\partial_{x_i}^2\frac{\delta}{i\delta j^{a_i\mu_i}(x_i)}\Bigg)\,.
\end{align} 
We can then formally calculate the contribution to an observable by squaring, weighting the matrix element, and summing over all such amplitudes:
\begin{align}
\langle\hat{O}\rangle_{N}&=\sum_{n=0}^{\infty}\int \prod_{i=1}^n[d^dq_i]_+\Big|\mathcal{A}_{N}^{[n]}(q_1^{a_1\lambda_1},...,q_n^{a_n\lambda_n})\Big|^2O^{a_1\lambda_1,...,a_n\lambda_n}(q_1,...,q_n)\,.
\end{align}
The function $O$ gives the final state contribution to some pre-defined observable. The factorization theorem for the hard scattering process then takes the form:
\begin{align}\label{eq:standard_factorization_theorem_N_Jet}
\frac{d\sigma}{dO}&=\text{tr}\Big[\mathbf{H}_N\langle\hat{O}\rangle_{N}\Big]\otimes_{i=1}^{N} J_i(\hat{O})\,,
\end{align}
where now the trace is over the color. $J_i$ here are the jet functions for the initial N hard partons and include contributions from collinear emissions(with energy of the same order as the N partons). Of particular importance are so-called additive observables. We will call an observable additive if it satisfies in \emph{some} conjugate space the property\footnote{We call this additive in deference to event shapes where this additivity is in the exponent of the Laplace/Fourier transform.}:
\begin{align}\label{eq:additive_observable}
O^{a_1\lambda_1,...,a_n\lambda_n}(q_1,...,q_n)&=O^{a_1\lambda_1}(q_1)O^{a_2\lambda_2}(q_2)....O^{a_n\lambda_n}(q_n)\,\qquad\forall n.
\end{align}
Note that the observable for our purposes only needs to be additive in the soft limit with respect to the hard scale of the eikonal lines. We have not taken the measurement to be independent of color and spin, but generally, one can relax this constraint and still have additivity. For most event shapes, one has
\begin{align}
O^{a\lambda}(q)&=O(q)\,.
\end{align}

Having introduced the LSZ reduction procedure to generate final states, we can give a formal definition to the insertion of a complete set of states connecting a soft amplitude functional and its conjugate weighted by a particular observable:
\begin{align}\label{eq:squaring_soft_amplitudes_for_an_observable}
\langle\hat{O}\rangle_{N}&=\sum_{n=0}^{\infty}\hat{O}^{[n]}\Big(\frac{\delta^{2}}{-i^2\delta j^+\cdot\delta j^-}\Big)\mathbf{Y}_N^{\dagger}[j^-]\mathbf{Y}_N[j^+]\Bigg|_{j^+=j^-=0}
\end{align}
Where we have defined the LSZ measurement operator for the $n$-parton contribution to the observable $\hat{O}$:
{\small\begin{align}\label{eq:LSZ_reduction_for_Soft_Functions}
\hat{O}^{[n]}\Big(\frac{\delta^{2}}{\delta j^+\cdot\delta j^-}\Big)&=\prod_{i=1}^n\int[d^dq_i]_+O^{a_1\lambda_1,...,a_n\lambda_n}(q_1,...,q_n)\nonumber\\
&\qquad\qquad \int d^dx_i\int d^dy_ie^{iq_i\cdot( x_i-y_i)}\epsilon_{q_i}^{\mu_i}(\lambda_i)\epsilon_{q_i}^{\nu_i}(-\lambda_i)\partial_{x_i}^2\partial_{y_i}^2\frac{\delta^2}{-i^2\delta j^{+a_i}_{\mu_i}(x_i)\delta j^{-a_i\nu_i}(y_i)}\\
&=\prod_{i=1}^n\int[d^dq_i]_+O^{a_1\lambda_1,...,a_n\lambda_n}(q_1,...,q_n)L\Big(q_{1}^{a_1\lambda_1},...,q_{n}^{a_n\lambda_n};\frac{\delta}{\delta j^+}\Big)L\Big(q_{1}^{a_1\lambda_1},...,q_{n}^{a_n\lambda_n};\frac{\delta}{\delta j^-}\Big)\,.
\end{align} }%
Formally, \Eq{eq:squaring_soft_amplitudes_for_an_observable} is nothing more than the insertion of a complete set of states, a method often used to calculate a soft function in SCET when there are no straightforward dispersion relations relating the soft function to a time-ordered product. Often such a insertion of a complete set of states is written as:
\begin{align}\label{eq:SCET_soft_function}
\langle\hat{O}\rangle_{N}&=\sum_{X_s}O(X_s)\langle 0|\bar{T}\{\mathbf{S}_{p_1}...\mathbf{S}_{p_n}\}|X_s\rangle\langle X_s|T\{\mathbf{S}_{p_1}...\mathbf{S}_{p_n}\}|0\rangle
\end{align}
The reason we adopt the more formal LSZ reduction approach to inserting states given in \Eq{eq:squaring_soft_amplitudes_for_an_observable} is that we can write an explicit integral equation (for example, \Eq{eq:integral_eq_for_phase_space}) accurate to a specific logarithmic order for the \emph{recursive} insertion of these states. 

We can of course rephrase the functional derivatives as a particular in-in path integral, see Refs. \cite{Schwinger:1960qe,Keldysh:1964ud}. We refer the reader to Ref. \cite{Belitsky:1998tc} for an illustration of the in-in path integral for a soft function calculation, see also Ref. \cite{Balitsky:1990ck}. For simplicity, we will work in pure Yang-Mills, but the results straightforwardly extend to full QCD. We write:
{\small\begin{align}
\langle\hat{O}\rangle_{N}=\int \mathcal{D}A^{+}\mathcal{D}A^{-} \mathbf{S}_{p_1}^{+}...\mathbf{S}_{p_n}^{+}\mathbf{S}_{p_1}^{-}...\mathbf{S}_{p_n}^{-}\text{Exp}\Big[iS[A^+]-iS[A^-]\Big]\prod_{x\in \Sigma_{\infty}}\delta(A^+(x)-A^-(x))\prod_{x\in \Sigma_{\infty}}\hat{O}(A^+(x),A^{-}(x)) 
\end{align}}
Here $\Sigma_\infty$ denotes a space-like surface at time $t=\infty$ upon which the $+$ and $-$ fields are constrained to be equal. The observable $\hat{O}$ then constrains the energy-momentum configuration on this surface, for instance by specifying the value of all suitably smeared C-correlators on this surface, as defined in Refs. \cite{Sveshnikov:1995vi,Cherzor:1997ak}. Within the in-in formalism, any free-theory two-point function connecting a $+$ to a $-$ field is equivalent to using an on-shell delta function for the propagator, and then we use time-ordered and anti-time ordered propagators to complete the basis for the total in-in propagator. Such $+,-$ two-point functions are explicitly realized in Eq. \eqref{eq:squaring_soft_amplitudes_for_an_observable} with the LSZ reduction procedure.

\subsection{Connecting SCET$_+$ and the Generating Functional}
Often one is interested in a hierarchy of soft scales, so that between the hard eikonal lines and the softest unresolved emissions are a set of observed softer jets. These soft jets could have hierarchies amongst themselves or not, but minimally there is a set of soft emissions (with respect to the hard initial jets) that are strongly ordered with respect to all other soft emissions (which are soft compared to all other scales, except perhaps $\Lambda_{\text{QCD}}$). An example of an observable which gives this type of hierarchy will be discussed in the next section.  We now write out the soft gluon matching for these ordered emissions in terms of the vacuum expectation value of wilson lines, and derive analogous results for the color-space generating functional.

Suppose we do not want to keep the full set of soft emissions (which are softer than the initial hard partons but otherwise arbitrary), but only want to retain the emissions in a (partially) strongly ordered limit. Specifically, if one does not set the current to zero in Eq. \eqref{eq:LSZ_reduction_Amplitude}, then one can order the momenta coupled to the current to be at a soft or collinear scale lower than the explicit momentum insertions enforced by the LSZ operator. This leads to a matching equation that can be derived following arguments of Refs. \cite{Bauer:2011uc,Larkoski:2015zka,Pietrulewicz:2016nwo}:
\begin{align}\label{eq:LSZ_reduction_matching}
&L\Big(q_{1}^{g_1\sigma_1},...,q_{n}^{g_n\sigma_n};\frac{\delta}{\delta j}\Big)\mathbf{Y}_N^{a_1b_1,...,a_Nb_N}[j,\mu]\prod_{k=1}^NJ_{n_k}^{b_k\nu_k}(\omega_k^{c_k\lambda_k};\mu)=\nonumber\\
&\qquad \mathbf{J}_{N}^{a_1b_1,...,a_Nb_N}(q_1^{\mu_1 e_1},...,q_n^{\mu_n e_n},\mu)\mathbf{Y}_{N+n}^{b_1d_1,...,b_Nd_N,e_1f_1,....,e_nf_n}[j,\mu]\prod_{k=1}^NJ_{n_k}^{d_k\nu_k}(\omega_k^{c_k\lambda_k};\mu)\prod_{i=1}^nJ^{\mu_if_i}_{n_i}(\omega_i^{g_i\sigma_i};\mu)+...\,.
\end{align}
We have added an additional $n$ wilson lines to the original soft function, and color contracted into the new hard matching coefficient. We have also made all color ($a_k,b_k,c_k,d_k,e_i,f_i,g_i$) and spin ($\nu_k,\lambda_k,\mu_i,\sigma_i$) indices explicit in the soft and jet functions, at risk of cluttering the equation. We have not necessarily ordered any of the emissions $q_i$ amongst themselves, only with respect to additional emissions beyond $q_i$, $i=1,...,n$. It is clear that one could then recursively apply such soft factorizations, which is equivalent to an ordering assumption on various subsets of the final state momenta. Note that in this equation, we have explicitly kept track of the color flow and polarization dependence of the new field insertions, but left the color indices of the original eikonal lines unspecified. The $\mathbf{J}_N$ are the matching coefficients for the factorization, and will in general be related to soft currents familiar from QCD factorization. Indeed, this must be true, since after setting $j=0$, we must reproduce Eq. \eqref{eq:LSZ_reduction_Amplitude}. The matrix elements become scaleless, and would feed the collinear polarizations and color indices to the currents, forming the appropriate soft amplitudes. We have also introduced collinear matrix elements:\footnote{For conciseness, we have written the operator implicitly as if they were all gluons, using a Lorentz vector index $\mu$ for the fields. Technically, we can also have other collinear field operators for fermions or scalars.}
\begin{align}\label{eq:jet_amplitudes}
J^{\mu_i c_i}_{n_i}(\omega_i^{a_i\lambda_i})=\langle 0|\delta(\nbar_i\cdot q_i-\bar{\mathcal{P}}_i)X_{n_i}^{c_i\mu_i}(0)|q_{i}^{a_i\,\lambda_i}\rangle
\end{align}
for each of the new soft eikonal lines, and $n_i,\bar{n}_i$ form a light-cone basis for the i-th direction. We also emphasize this matching equation holds for \emph{renormalized} quantities on both sides of the equation. 

The color-space generating functionals also satisfy an analogous factorization, now expressed in terms the generalization of the color space currents. Ignoring the particular loop order the current has been evaluated to, we may write:
\begin{align}\label{eq:color-space_generating_functional_factorization}
\mathbf{J}^{[n]}(U)\mathcal{W}_{N}\Big(U\Big)&=\int\prod_{i=1}^n[d^dq_i]_+\mathcal{A}_N^{[n]}\cdot\mathcal{W}_{N+n}\Big(U\Big)\,,
\end{align} 
where one recalls that the color-space generating functional was defined with specific directions for the hard eikonal directions. In particular, we can then identify the vacuum expectation value of the $N$-eikonal line soft matrix element with the generating functional for soft amplitudes:
\begin{align}\label{eq:color_space_to_soft_wilson_lines}
\mathcal{A}_N^{[n]}&\leftrightarrow   \mathbf{J}_{N}(q_1^{\mu_1 b_1},...,q_n^{\mu_n b_n},\mu)\\
\mathcal{W}_{N}\Big(U\Big)&\leftrightarrow \langle 0|T\Big\{\mathbf{S}_{p_1}...\mathbf{S}_{p_N}e^{iS[A]+i j\cdot A}\Big\}|0\rangle\prod_{k=1}^NJ_{n_k}(\omega_k)
\end{align}
We have suppressed the color and spin indices for conciseness. Finally, we have the basic connection between $U$-averaged generating functional in Eq. \eqref{eq:soft_trace_gen_func} and the insertion of a complete set of states in an SCET soft function \eqref{eq:SCET_soft_function}. In general, we must consider both the soft and collinear contributions since the soft anomalous dimension depends on the large momentum of each collinear sector. We can define the ordered infra-red functions:\footnote{For the time being we ignore PDFs.}
\begin{align}\label{eq:matrix_element_definition_parton_shower_state}
\mathbf{S}_N(\hat{O},\mu_H,\mu)&=\sum_{|X_{s,c}|<\mu_H}O\Big(X_s,X_c;\{p_i\}_{i=1}^N\Big)\langle 0|\bar{T}\{\mathbf{S}_{p_1}...\mathbf{S}_{p_N}\}|X_s\rangle\langle X_s|T\{\mathbf{S}_{p_1}...\mathbf{S}_{p_N}\}|0\rangle\prod_{k=1}^{N}J_{p_k}(X_c)\\
J_{p_k}(X_c;\mu)&=N_k\text{tr}\langle 0|X_{p_k}(0)\delta(\bar{n}_k\cdot p_{k}-\bar{n}_k\cdot\mathbb{P})\delta_{\perp p_k}^{(2)}\Big(\mathbb{\vec{P}}_\perp\Big)|X_c\rangle\langle X_c|X_{p_k}(0)|0\rangle
\end{align} 
Where we use collinear field operators defined in Eq. \eqref{eq:collinear_operators}. $\mathbb{P}$ denotes the momentum operator, and the subscript $\perp p_{k}$ denotes the projection onto the plane transverse to the null direction $p_k$. The scale $\mu$ denotes where the ultra-violet virtual divergences are renormalized, and $|X_{s,c}|<\mu_H$ denotes that the final soft $X_s$ states and the collinear $X_c$ states are to be ordered together. For us, the ordering of the real emissions is dictated by the ordering of the soft anomalous dimension as derived in the effective theory in App. \ref{app:soft_virt}: the on-shell region of the virtual corrections must match the real emissions, so that we are guaranteed no spurious large logarithms and the correct cancellation of infra-red divergences.\footnote{In the large-$N_c$ limit, this argument is essentially an unitarity argument, as the soft anomalous dimension gives the no splitting probability, and the splitting phase-space must according match.} Then the resummation of $\mathbf{G}_N$ as asserted in Eq. \eqref{eq:leading_log_gen_functional} follows from using the observable constrained sum over states found in \eqref{eq:LSZ_reduction_for_Soft_Functions}, the factorization properties found in Eq. \eqref{eq:LSZ_reduction_matching} for the case of a single additional emission, which drives the connection between $\mathbf{G}_N$ and $\mathbf{G}_{N+1}$ within the BMS equation.

We have made no assumption about the factorization properties of the measurement. $N_k$ is the appropriate normalization factor for the jet function, so its tree level value is $1$ for a cumulative distribution, given we trace over color and spin indices for the jet function. We emphasize that Eq. \eqref{eq:matrix_element_definition_parton_shower_state} must be defined with overlaps between the soft and collinear states removed.\footnote{In the parton shower literature, this would be called \emph{sectorizing} the shower. In the EFT language, we must perform a zero-bin subtraction.} 

We should also clarify the recoil constraints. Each momentum $p_K$ used to define a collinear sector can be required to satisfy exact momentum conservation through the measurement constraint $O\Big(X_s,X_c;\{p_i\}_{i=1}^N\Big)$. If we define each jet function relative to a recoil sensitive jet axis, for example the thrust axis, the final state of the jet function will not necessarily align with $p_k$, and we could induce an explicit transverse momentum injected into the jet function. 

Then we have:
\begin{align}\label{eq:pQCD_and_SCET}
\mathbf{G}_N(O,\mu_H)&=\mathcal{U}_{N}(\mu_H,\mu)\mathbf{S}_N(\hat{O},\mu_H,\mu)\mathcal{U}_{N}^{\dagger}(\mu_H,\mu)\,.
\end{align}
We note that this construction is formally independent of $\mu$, where we renormalize the virtual corrections. This relationship will hold so long as we insist that the measurement constraints in the $U$-averaging found in Eq. \eqref{eq:ave_with_obs} are identical to the constraints used in $O\Big(X_s,X_c;\{p_i\}_{i=1}^N\Big)$. This object will evolve with Eq. \eqref{eq:BMS_equation_V1} or equivalently Eq. \eqref{eq:BMS_for_dressed_gluons} once we use transverse-ordering. In the language of SCET, this is essentially saying that we evolve our hard function(which is the same as the virtual corrections) from the hard scale to some IR scale $\mu$ at which the soft and jet functions are evaluated, making the whole object $\mathbf{G}_N$, $\mu$ independent.

For a specific observable, we can often make use of the multipole expansion to significantly simplify Eq. \eqref{eq:matrix_element_definition_parton_shower_state}, disentangling the collinear and soft sectors. However, if we are not careful in the power counting of the modes used for the multipole expansion, all large logs in phase-space may not be resummed by the resulting multipole expanded functions. We now elaborate on this point.
 
\subsection{Constructing Evolution Equations for a Multipole Expanded Measurement}
We now examine constructing a BMS equation with a strict multipole expansion using a hard-soft factorization, as in as Ref. \cite{Becher:2016omr}, for the case of hemisphere jet masses. We ignore the transverse ordering that the Glauber mediated interactions impose. We will adopt a theory with two modes:
\begin{align}\label{eq:naive_power_count}
p_H&\sim m_H(1,1,1)\,,\nonumber\\
p_S&\sim m_L(1,1,1)\,.
\end{align}
Ref. \cite{Becher:2016omr} argued that collinear modes need not be introduced, and we will examine the consequences of such a power counting. Indeed, we will see such a theory is perfectly consistent, but does not resum all large logarithms of \emph{phase-space}. That is, it will miss multiple emission effects near the boundary of the two hemispheres, or at and below the scale $m_L$. Evolving with the object defined in Eq. \eqref{eq:pQCD_and_SCET} with Eq. \eqref{eq:BMS_equation_V1} or equivalently Eq. \eqref{eq:BMS_for_dressed_gluons} will not suffer from such a limitation, since one would use the correct phase-space constraints for all emissions (not presupposing a multipole expansion of the phase-space), populating formally all emissions down to $\mu_S=0$, or more appropriately, the hadronization scale. Put simply, the full parton-shower will populate both hemispheres with multiple emissions as the phase-space constraints allow.

\subsubsection{Structure of Observables}
Before deriving the BMS master equation, we first clarify the structure of the additive observable. In the wilson line approach, the LSZ reduction procedure allows one to construct all the necessary scattering amplitudes, so to form the soft expectation value, one simply needs to stitch together the appropriate amplitudes, as is done in Eq. \eqref{eq:LSZ_reduction_for_Soft_Functions}. We take as an illustrative example the non-global observable of a fat jet cumulative energy($m_H$), and the complement region's cumulative energy($m_L$) such that $m_H>m_L$.  For $n$ final state partons, this is the LSZ measurement operator:
{\small\begin{align}\label{eq:N_emission_jet_energy_obs}
\hat{\mathcal{I}}&_{m_H,m_L}^{[n]}=\nonumber\\
&\prod_{i=1}^n\int[d^dq_i]_+\theta\Big(m_H-\sum_{i=1}^nn\cdot q_{i}\theta_{J}(q_i)\Big)\theta\Big(m_L-\sum_{i=1}^n\nbar\cdot q_{i}\theta_{\overline{J}}(q_i)\Big) L\Big(q_{1}^{a_1\lambda_1},...,q_{n}^{a_n\lambda_n};\frac{\delta}{\delta j^+}\Big)L\Big(q_{1}^{a_1\lambda_1},...,q_{n}^{a_n\lambda_n};\frac{\delta}{\delta j^-}\Big)\nonumber\\ 
\end{align}}%
Note that we do not apply any momentum conservation constraints on this operator, this is in accord with the SCET multipole expansion for hemisphere soft-functions probing the jet mass, see Ref. \cite{Fleming:2007qr}. The observable itself to all orders is then:
\begin{align}\label{eq:all_emission_jet_energy_obs}
\hat{\mathcal{I}}_{m_H,m_L}=1+\sum_{n=1}^{\infty}\hat{\mathcal{I}}_{m_H,m_L}^{[n]}
\end{align}
An important consideration for any master equation is to find the correct ordering of the above phase space accurate to the logarithmic order one is working. In particular, one is interested in the region between $m_H$ and $m_L$. In order to trace over a particular emission (replacing it with a wilson line and a jet function), this emission must be above the softest scale $m_L$ in the observable. We introduce then the low-scale measurement operator:
{\small\begin{align}\label{eq:low_scale_emission_jet_energy_obs}
\hat{\mathcal{S}}_{m_L}^{[n]}&=\prod_{i=1}^n\int[d^dq_i]_+\theta\Big(m_L-\sum_{i=1}^n\nbar\cdot q_{i}\theta_{\overline{J}}(q_i)\Big) L\Big(q_{1}^{a_1\lambda_1},...,q_{n}^{a_n\lambda_n};\frac{\delta}{\delta j^+}\Big)L\Big(q_{1}^{a_1\lambda_1},...,q_{n}^{a_n\lambda_n};\frac{\delta}{\delta j^-}\Big)\nonumber\\
\hat{\mathcal{S}}_{m_L}&=1+\sum_{n=1}^{\infty}\hat{\mathcal{S}}_{m_L}^{[n]}
\end{align}}%
The integral is unrestricted when the soft emission is in the fat jet region, due to the multipole expansion, see Ref. \cite{Becher:2016mmh}. Then we have the ordered (resolved) emissions measurement operator:
{\small\begin{align}\label{eq:resolved_emission_jet_energy_obs}
\hat{\mathcal{R}}_{m_H,m_L}^{[n]}&=\prod_{i=1}^n\int[d^dq_i\theta_{J}(q_i)]_+\theta\Big(m_H-\sum_{i=1}^nn\cdot q_{i}\Big)\theta(n\cdot q_1-n\cdot q_2)\nonumber\\
&\qquad\times\theta(n\cdot q_2-n\cdot q_3)...\theta(n\cdot q_{n-1}-n\cdot q_n)\theta(n\cdot q_n-m_L)\nonumber\\
&\qquad\qquad L\Big(q_{1}^{a_1\lambda_1},...,q_{n}^{a_n\lambda_n};\frac{\delta}{\delta j^+}\Big)L\Big(q_{1}^{a_1\lambda_1},...,q_{n}^{a_n\lambda_n};\frac{\delta}{\delta j^-}\Big)\nonumber\\
\hat{\mathcal{R}}_{m_H,m_L}&=1+\sum_{n=1}^{\infty}\hat{\mathcal{R}}_{m_H,m_L}^{[n]}
\end{align}}
The ordering involves no approximations, at least in a pure yang-mills theory\footnote{And also for a theory where all flavors have the same charge representation. Thus they give rise to ``identical'' Wilson lines. One still would have to consider the jet function contributions.}, since the matrix element is invariant under interchanges, so that one can impose and rearrange an ordering to a canonical form. The major approximation is that the softest emission is still above the scale $m_L$, so that all emissions are at the fat jet scale. Note that the emissions must be in the fat jet region, given they are all above the scale $m_L$. 

Now we can write down the leading logarithmic recursion relation that can be made to the resolved emission measurement operator\footnote{This leading logarithmic expansion of the phase space is the same as that done in the coherent branching algorithm of \Ref{Catani:1990rr}. Indeed, there they showed the leading logarithmic expansion of the phase space works to next-to-leading log as well. Beyond leading log order, one would use the same ordering and factorization between low and high-scale measurements, but in laplace space variables.}:
\begin{align}
\hat{\mathcal{R}}_{m_H,m_L}^{[n]}&=_{LL}\int[d^dq\theta_{J}(q)]_+\theta\Big(m_H-n\cdot q\Big)\hat{\mathcal{R}}_{n\cdot q,m_L}^{[n-1]}L\Big(q^{a\lambda};\frac{\delta}{\delta j^+}\Big)L\Big(q^{a\lambda};\frac{\delta}{\delta j^-}\Big)
\end{align}
This can be solved using energy ordered exponentials, defining the argument of the exponential to be:
\begin{align}
\mathcal{E}(\lambda;\delta_{j^+}\cdot\delta_{j^-})&=\int[d^dq\theta_{J}(q)]_+\delta(\lambda-n\cdot q) L\Big(q^{a\lambda};\frac{\delta}{\delta j^+}\Big)L\Big(q^{a\lambda};\frac{\delta}{\delta j^-}\Big)\\
\hat{\mathcal{R}}_{m_H,m_L}^{LL}&=T_E\text{ Exp}\Bigg[\int_{m_L}^{m_H} d\lambda'\,\mathcal{E}(\lambda';\delta_{j^+}\cdot\delta_{j^-})\Bigg]\label{eq:resolved_emission_jet_energy_obs_LL}
\end{align}
This last relation follows from using the fact:
\begin{align}
\int_{\lambda_i}^{\lambda_f}d\lambda'\delta(\lambda'-\lambda)&=\theta(\lambda_f-\lambda)\theta(\lambda-\lambda_i)
\end{align}
Often, we will find it useful to consider the integral equation satisfied by the resolved emission measurement operator:
\begin{align}\label{eq:integral_eq_for_phase_space}
\hat{\mathcal{R}}_{m_H,m_L}^{LL}&=1+\int_{m_L}^{m_H}d\lambda\,\hat{\mathcal{R}}_{\lambda,m_L}^{LL}\mathcal{E}(\lambda;\delta_{j^+}\cdot\delta_{j^-})
\end{align}
Finally, for the purpose of computing non-global distributions, we need the full measurement operator to factorize at leading power in $m_L/m_H$:
\begin{align}\label{eq:all_emission_jet_energy_obs_factorization}
\hat{\mathcal{I}}_{m_H,m_L}=\hat{\mathcal{S}}_{m_L}\hat{\mathcal{R}}_{m_H,m_L}+O\Big(\frac{m_L}{m_H}\Big)
\end{align}
With Eq. \eqref{eq:all_emission_jet_energy_obs_factorization}, we act:
\begin{align}\label{eq:LSZ_reduction_for_Soft_Functions_additive}
\Big\langle\hat{\mathcal{I}}_{m_H,m_L}\Big\rangle_{N}&=\hat{\mathcal{I}}_{m_H,m_L}\mathbf{Y}_N^{\dagger}[j^{+}]\mathbf{Y}_N[j^-]\Big|_{j^+=j^-=0}\,.
\end{align}
Then after we make use of the integral equation \eqref{eq:integral_eq_for_phase_space} with the factorization in Eq. \eqref{eq:all_emission_jet_energy_obs_factorization}, and then renormalize all global divergences, that is, divergences associated with becoming parallel to either the $n$ or $\nbar$ directions as defined by the thrust axis, we can derive a minimal BMS equation to resum hemisphere distributions. The critical point is the infra-red boundary condition, which is given by the soft matrix elements:
\begin{align}\label{eq:scet_boundary}
\Big\langle\hat{\mathcal{S}}_{m_L}\Big\rangle_{N}\,.
\end{align}
We now turn to the one-loop expression for these matrix elements.

\subsubsection{Limits of the Hard-Soft Factorization}\label{eq:Beyond_Everything}
So far, we have only considered the BMS equation in a factorization between only hard modes and soft modes. One might think this is justified for observables dominated by soft physics, the non-global observables. One can straightforwardly factor out any collinear effects associated with the Sudakov double logarithms, and the left over result will be a single logarithmic series. The issue encountered is two fold, but related: the kernels for multiple soft emissions that make up the BMS equation at higher orders are not uniform in phase-space, but have certain regions that are enhanced due to collinear splittings \cite{Hatta:2017fwr}, and the boundary conditions in the infra-red have enhanced regions of phase space due to collinear splittings \cite{Neill:2015nya,Larkoski:2016zzc}. Here we will focus on the latter case, merely pointing out which logarithms in the fixed order calculation of the boundary condition are problematic. The important point to emphasize, though, is that these large logarithms are integrable order-by-order in perturbation theory, since infra-red and collinear divergences cancel in the BMS equation. One can work to a formally well-defined order in the soft-logarithms without resumming collinear effects. However, large logarithms integrate to large constants, so the question arises whether this formally well-defined order is genuinely perturbatively stable, i.e., the next correction is smaller than what has been retained.

What is an example of a large logarithm in the infra-red boundary condition? We take the case of hemisphere dijet invariant mass spectra in $e^+e^-$ collisions as an example. The result for the boundary condition was calculated in Refs. \cite{Becher:2016mmh,Becher:2016omr}, but also can be extracted from the results of Ref. \cite{Ellis:2010rwa}, due to their decomposition of phase space integral for a single soft emission off of an arbitrary set of hard jets. Let $n$ be the null vector defined by the thrust axis pointing into the heavy hemisphere, and let $\theta_i$ be the angle the emission $i$ generated in the course of the BMS evolution makes to the thrust axis. Finally $\phi_i$ is the azimuthal angle that emission $i$ has in the transverse plane dividing the hemispheres. Then we have for the boundary condition (in laplace-space):
{\small\begin{align}
\int_{0}^{\infty} dm_L e^{-\tau_L m_L}\Big\langle\hat{\mathcal{S}}_{m_L}[\mu]\Big\rangle_{N}\Big|_{one-loop}&=\sum_{i,i\neq n}\mathbf{T}_n\cdot \mathbf{T}_i\frac{\alpha_s(\mu)}{4\pi} u(\tau_L\mu,\theta_{i})+\sum_{i,j,i\neq n,j\neq n}\mathbf{T}_i\cdot \mathbf{T}_j\frac{\alpha_s(\mu)}{8\pi} v(\tau_L\mu;\theta_i,\theta_j,\phi_i-\phi_j)\,.
\end{align}}
Let us single out a particular emission $i$, so that we can write:
{\small\begin{align}
\int_{0}^{\infty} &dm_L e^{-\tau_L m_L}\Big\langle\hat{\mathcal{S}}_{m_L}[\mu]\Big\rangle_{N}\Big|_{one-loop}=\nonumber\\
&\mathbf{T}_n\cdot \mathbf{T}_i\frac{\alpha_s(\mu)}{4\pi} u(\tau_L\mu,\theta_{i})+\sum_{j\neq n}\mathbf{T}_i\cdot \mathbf{T}_j\frac{\alpha_s(\mu)}{8\pi} \Big(v(\tau_L\mu;\theta_i,\theta_j,\phi_i-\phi_j)+v(\tau_L\mu;\theta_j,\theta_i,-\phi_i+\phi_j)\Big)+...\,.
\end{align}}
Where the ``$...$'' refers to terms not involving the eikonal line $i$. Importantly, as $\theta_{i}\rightarrow \frac{\pi}{2}$:
{\small\begin{align}
\int_{0}^{\infty} dm_L e^{-\tau_L m_L}&\Big\langle\hat{\mathcal{S}}_{m_L}[\mu]\Big\rangle_{N}\Big|_{one-loop}=\nonumber\\
&\frac{\alpha_s(\mu)}{4\pi} \Big(\mathbf{T}_n\cdot \mathbf{T}_i+\sum_{j\neq n}\mathbf{T}_i\cdot \mathbf{T}_j\Big)\Big(-4\text{ln}(\mu\tau_L)\text{ln}\Big(1-\text{tan}^2\frac{\theta_i}{2}\Big)+2\text{ln}^2\Big(1-\text{tan}^2\frac{\theta_i}{2}\Big)\Big)+...\,,\\
&=-\frac{\alpha_s(\mu)}{2\pi}\mathbf{T}_i^2\Big(-2\text{ln}(\mu\tau_L)\text{ln}\Big(1-\text{tan}^2\frac{\theta_i}{2}\Big)+\text{ln}^2\Big(1-\text{tan}^2\frac{\theta_i}{2}\Big)\Big)+...
\end{align}}
This is precisely the structure of logarithms found in Ref. \cite{Neill:2015nya}, where a \emph{jet} function was introduced to resum the large phase-space logarithm $\text{ln}\Big(1-\text{tan}^2\frac{\theta_i}{2}\Big)$, called the ``edge-of-jet'' function. These jet functions of course can be introduced directly into the BMS equation itself. That they should appear is clear from Eq. \eqref{eq:LSZ_reduction_matching}, since it contains collinear functions. One then forms jet functions once one squares the SCET$_+$ factorization, with a phase space sensitive to the boundary of the active jet region, not expanding their phase-space as if they are deep inside the active jet region (that gives scaleless integrals for the jet function). Physically, the appearance of these logarithms in the boundary conditions corresponds to a situation where as a jet populated by the BMS evolution approaches the boundary of the active jet region, it may collinearly split in and out of the active region. Then the boundary of the active region acts as a collinear cut-off which will not cancel out due to the mismatch in the scales between the light and heavy hemispheres. These logarithms are integrable, but give rise to large contributions to the coefficients of the fixed-order expansion for the non-global logarithms, as detailed in Ref. \cite{Larkoski:2016zzc}. The leading log series will have a finite radius of convergence, while the collinear splittings eventually will act as a sort of renormalon as one pushes to higher orders in the BMS resummation, unless the collinear splittings themselves are resummed. Merely adopting the renormalization scale $\mu\sim m_L\sim \tau_{L}^{-1}$, the scale choice consistent with a simple decomposition into only soft and hard regions dominate as advocated in Ref. \cite{Becher:2016mmh} does not appear to suffice to control all large logarithms in phase-space. That is, it appears we must work beyond the soft approximation to have a controlled perturbation series.\footnote{Of course, given the complexity of the BMS equation, it is important to have all the simplifying approximations one can make in order to understand it. Some form of the multipole expansion has therefore been used in Refs. \cite{Dasgupta:2001sh,Banfi:2002hw,Weigert:2003mm,Caron-Huot:2015bja,Larkoski:2015zka,Neill:2016stq}.}

We contrast this state of affairs with the evolution with the transverse-ordered BMS equation encoded in Eq. \eqref{eq:BMS_for_dressed_gluons} (which we may take to act on effective field theory objects as in \eqref{eq:pQCD_and_SCET}). There we do not multipole expand the phase space for each emission, keeping the constraints exact, allowing the evolution equation to populate the phase space as the measurement constraints allow. This is exactly analogous to how the transverse-ordered BMS equation reproduced the NLL resummation for global observables as accomplished, eventually resumming the large logarithms with explicit population of multiple emissions at the infra-red inclusive scales. Evolving the transverse-ordered BMS equation down to zero cutoff (or the hadronization scale), formally below the completely inclusive scale $m_L$, would resum the large phase-space logarithms in the low-scale soft function $\Big\langle\hat{\mathcal{S}}_{m_L}\Big\rangle_{N}$ that acts as the infra-red boundary condition for the multipole expanded BMS equation, using the strict hard-soft factorization without collinear modes.


\section{Conclusions}

We have presented two ways to understanding how soft radiation in a non-abelian gauge theory dresses a hard interaction, both leading to the BMS equation. The first way introduced a generating functional built from the soft anomalous dimension. This anomalous dimension controls the infra-red divergences of exclusive hard scattering coefficients at the level of the S-matrix amplitude. Real-emissions are handled with the eikonal currents for real emissions, following Refs. \cite{Weigert:2003mm,Caron-Huot:2015bja}. These currents are related to the factorization properties of wilson lines matrix elements with external states \cite{Bauer:2011uc,Larkoski:2015zka,Pietrulewicz:2016nwo}, and the anomalous dimension to the renormalization of the wilson lines. We then see that the EFT operators deformed in the presence of an external current really are nothing other than the generating functional built from real and virtual diagrams. The BMS equation itself can be seen as a way of organizing the Fock space of resolved emissions. Iterating the BMS equation while exponentiating all virtual corrections produces the dressed gluon/out-of-gap expansion of Refs. \cite{Forshaw:2006fk,Forshaw:2008cq,Forshaw:2012bi,Larkoski:2015zka,Larkoski:2016zzc}, which can been seen as equivalent to taking the trace of a reduced density matrix of all resolved emissions against the appropriate operator specifying the observable.

The most important new result is a straightforward proof of the out-of-gap/dressed gluon expansion of Ref. \cite{Angeles-Martinez:2015rna} when both initial state and final state eikonal lines are required. Once one exposes the Cheshire Glauber in the soft region with a zero-bin subtraction, the transverse ordering of the soft anomalous dimension is the only sensible way to reproduce the correct imaginary part, if one wants to trivialize the zero-bin. The Glauber region produces the imaginary part of the soft anomalous dimension, and the Glauber region gets its ultra-violet divergence from integrating over the transverse momentum of exchanged between a pair of partons. Within the multipole expanded Glauber potential of the EFT, this is the only foliation of momentum space that induces ultra-violet divergences, without interfering with the light-cone integrals and the resulting $i\pi$. Then the BMS equation with both initial and final state lines, with the appropriate ordering variable, can be seen as resumming all the factorization violating effects associated with the hard process in a hadron-hadron collision driven by active-active Glauber exchanges.

However, the transversely ordered BMS equation does not itself constitute a complete solution to the underlying event or factorization violation. It is still the evolution equation for an effective theory of semi-infinite wilson lines and perhaps additional collinear splittings. Thus it cannot describe to all orders spectator-spectator interactions, where Glauber exchanges do not eikonalize. Given a set of eikonal lines, it shows how to populate another \emph{final state} wilson line, which will contribute to the observable. To describe spectator-spectator interactions and thus completely factorize the underlying event one would have to create in the course of BMS evolution multiple additional initial-state (spectator) eikonal lines also participating in the hard interaction\footnote{Here we do not necessarily mean \emph{the} hard interaction (like the Higgs production region), but any energy constraint or veto placed on the final state above the scale $\Lambda_{QCD}$ (like a $p_T$ veto or gaps-between-jets).}. But accomplishing that goal would seem to require something like the B-JIMWLK \cite{Balitsky:1995ub, Kovchegov:1999yj,JalilianMarian:1996xn,JalilianMarian:1997gr,Iancu:2001ad} hierarchy to describe the initial state, interfaced with the BMS hierarchy to describe the production of additional jets. 

With regards the multipole expansion, we should stress that nothing wrong has occurred, only that simplifying assumptions about the momentum regions to be used can lead to a multipole expansion that induces large logs in certain regions of phase-space that are not resummed, even though the EFT is perfectly consistent with those modes. Moreover, for the effective theory formulation of the Lagrangian mediating all interactions, we have kept the multipole expansion and used the zero-bin subtraction for soft, collinear, and Glauber interactions for the construction of the soft anomalous dimension and all amplitude constructions, using SCET$_{II}$ power counting (Refs. \cite{Chiu:2009mg,Becher:2010tm,Chiu:2012ir}). We have only abandoned the multipole expansion for the phase-space. We can resum all logs using the multipole expansion with sufficient infra-red regions, leading to a more complicated BMS equation/dressed-soft-jet expansion including a jet function contribution and its resummation, or we can adopt exact momentum conservation in the shower, and evolve down to arbitrarily low scales (or till hadronization), using the measurement constraints on the phase-space as the regulator rather than dimensional/analytic regularization. Both will resum all large logs, however, the price we pay in the use of exact momentum conservation will be in-homogeneous power counting, obscuring the sub-leading power form of the shower. This seems a small price to pay given the complexity of the resummation of soft physics.

Thus the important conclusion is that we should really consider the full soft and collinear production after a hard interaction, even if we beguile ourselves into thinking we only need a hard/soft factorization. Non-global effects make it very difficult to make all orders statements. Indeed, transverse ordering of the shower only works so long as we consider the one real emission term of the BMS equation. At higher perturbative orders for the BMS equation kernel, there will be multiple emissions terms that are not strongly ordered interacting with multiple wilson lines in a non-dipole form. How these terms should be ordered is an open question, particular beyond leading color, and would require detailed calculations, though an energy-energy correlator like scheme similar to Refs. \cite{Larkoski:2013eya,Larkoski:2014gra,Larkoski:2016zzc} was adopted in Ref. \cite{Caron-Huot:2016tzz} for the BMS equation at large-$N_c$. Such observables form a natural extension of transverse-momentum for multiple emissions. Indeed, one may need to formulate a distinct ordering variable for each multiple emission term in the BMS equation.

In the large $N_c$ limit, always considering the full soft and collinear contributions is performed by almost all modern parton-showers, using full soft and collinear coherent matrix elements, though by taking the large $N_c$ limit they lose the factorization violation effects formally included here. Even in the large $N_c$ limit, adopting transverse ordering has long been understood as the best choice to resum the most effects in the parton shower. Still, considering the hard/soft factorization with the most aggressive multipole expansion of the traditional BMS equation is fruitful if for no other reason than it is simpler. But turning to the full parton shower, perhaps the most interesting theoretical questions to ask are: firstly, does the full parton shower accomplish the suggested collinear improvements to the BMS equation found in Ref. \cite{Hatta:2017fwr} and how? Secondly, what are the ``late-time'' asymptotics of the full parton shower, with exact momentum conservation? That is, along the lines of results in \cite{Neill:2016stq}, what are the changes in the structure of the buffer region \cite{Dasgupta:2002bw}, what is the diffusive behavior of the parton shower in active regions, how do Glauber gluon exchanges modify these asymptotics, and what are the impacts of collinear splittings and recoil, even in soft sensitive observables?

\section{Acknowledgments}
We would like to thank Ian Moult for reading the manuscript. This work was supported by the U.S. Department of Energy through the Office of Science, Office of Nuclear Physics under Contract DE-AC52-06NA25396 and by an Early Career Research Award, as well as through the LANL/LDRD Program.

\appendix

\section{Soft Virtual Corrections}\label{app:soft_virt}
Here we show how to justify the transverse ordering in the resummation of virtual corrections. We split the divergent parts of the soft integral determining the one-loop soft anomalous dimension into an ``on-shell'' term and an ``off-shell'' term. The ``on-shell'' term exactly matches the real emission eikonal factor by construction, and the ``off-shell'' component is the contribution from the Glauber exchange. Within the framework of soft-collinear effective field theory, the on-shell component corresponds to the naive soft contribution minus the Glauber region of the naive soft integral. The correct off-shell component will arise from insertions of the Glauber Lagrangian of Ref. \cite{Rothstein:2016bsq}. Of course, since we are dealing with active-active Glauber exchanges, by setting the directions of the wilson line appropriately, we need not formally consider the Glauber region at all, but leaving the Glauber region within the soft sector obscures its renormalization and how to justify transverse ordering. We have decompose the soft virtual emission into light-cone coordinates defined by two collinear sectors $i$ and $j$ as:
{\begin{align}
q^{\mu}&=\frac{1}{p_i\cdot p_j}\Big(p_i\cdot q\, p_j^{\mu}+p_j\cdot q\, p_i^{\mu}\Big)+q_{\perp ij}^{\mu}\,.
\end{align}}
The soft integral we want to consider is:
{\small\begin{align}
S_{ij}&=\int\frac{d^dq}{(2\pi)^{d}}\frac{p_i\cdot p_j}{(p_i\cdot q+i0)(q\cdot p_j+i0)}\times\frac{1}{q^2+i0}\,,\\
\label{eq:soft_loop_decomp}     &=\int[d^dq]_+\frac{p_i\cdot p_j}{(p_i\cdot q)(q\cdot p_j)}-i\lambda_{ij}\pi\int_0^{\infty}\frac{dq_{\perp ij}}{q_{\perp ij}}+...\,.
\end{align}}
The first term is the Glauber-bin subtracted soft integral, and the second is the explicit Glauber contribution, where $ \lambda_{ij} = 1$ if both $i,j$  are incoming, $0$ otherwise. If one explicitly calculates the Glauber Lagrangian contribution, the Glauber exchange between sectors $i$ and $j$ contributes as:
\begin{align}
\int\frac{dp_i\cdot q}{p_i\cdot q+i0}\frac{dp_j\cdot q}{p_j\cdot q+i0}\int\frac{d^{d-2}q}{-q_{\perp ij}^2}\rightarrow -i\pi\lambda_{ij}\int\frac{d^{d-2}q}{-q_{\perp ij}^2}\,.
\end{align}
This form of the Glauber contribution is necessary consequence of the multi-pole expansion of the effective theory, and the $i\pi$ is the necessary prescription for the rapidity divergent integrals as argued in Ref. \cite{Rothstein:2016bsq}. The ultra-violet divergences is manifestly in the transverse momentum alone, so the renormalization induces a running in the  scales.\footnote{If we choose some other cutoff, other than the transverse momentum, then one would mix the integration over the light-cone directions with the transverse momentum, and one would not produce the correct imaginary part in the rapidity regulated Glauber potential insertion.} If $q$ is on-shell, then:
\begin{align}
q^2=0\leftrightarrow \,\vec{q}_{\perp ij}^{\,2}=\frac{2 p_i\cdot q q\cdot p_j}{p_i\cdot p_j}
\end{align}
If we wish to impose an ultra-violet cutoff within the naive soft sector in Eq. \eqref{eq:soft_loop_decomp} that respects the form of the ultra-violet divergences from the Glauber exchange without interfering with the imaginary part of the naive soft sector, then we should break up the integral into contours of constant transverse momentum:
\begin{align}
S_{ij}(\mu)&=\int\frac{d^dq}{(2\pi)^{d}}\frac{p_i\cdot p_j}{(p_i\cdot q+i0)(q\cdot p_j+i0)}\times\frac{1}{q^2+i0}\mu\delta\Big(\mu-q_{\perp ij}\Big)\,.
\end{align}
We could of course attempt to slice the ultra-violet divergences in the naive soft sector however we wish, but then a non-trivial zero-bin would be necessary to insure we only renormalized the Glauber region contribution with respect to the transverse momentum integration, as dictated by the multipole expanded Glauber Lagrangian in the effective theory. This we avoid by imposing the transverse ordering on the entire naive soft sector. We can then forget about the zero-bin subtraction, since the zero-bin and the Glauber Lagrangian contributions will equal each other, and cancel. Splitting the soft integral into its on-shell (zero-bin subtracted) and Glauber-contributions we then have the result:
\begin{align}\label{eq:soft_integral_transverse_sliced}
S_{ij}(\mu)&=-i\lambda_{ij}\pi+\int[d^dq]_+\frac{p_i\cdot p_j}{(p_i\cdot q)(q\cdot p_j)}\mu\delta\Big(\mu-\sqrt{2W_{ij}^{-1}(q)}\Big)
\end{align}
This expression then sets the appropriate soft anomalous dimension including Glauber effects at both one and two loops in the perturbative expansion for soft anomalous dimension, given its dipole form to that order.

\subsection{On-shell Integral}

We evaluate \Eq{eq:soft_anom_dim_1_loop}, (for clarity, we denote the $\mu$ of the dimensional regularization procedure as $\bar\mu$, so that we are not necessarily identifying $\mu=\bar\mu$, though this will be done eventually):
\begin{align}\label{eq:soft_on-shell_virtual_integral}
\gamma_{ij}^{(1)}(\mu)&=\int[d^dq]_+\frac{p_i\cdot p_j}{p_i\cdot q\,q\cdot p_j}\mu\delta\Big(\mu-O(p_i,p_j;q)\Big)\nonumber\\
&=\frac{\bar \mu^{4-d}}{4\pi}\int_0^{\infty}\frac{d\omega}{\omega^{5-d}}\int\frac{d^{d-2}\Omega_{\hat{q}}}{(2\pi)^{d-2}}\frac{p_i\cdot p_j}{p_i\cdot n_q\,n_q\cdot p_j}\mu\delta\Big(\mu-O(p_i,p_j;q)\Big)
\end{align}
Where we have made use of the change of variables:
\begin{align}
q&=\omega \,n_q\\
n_q&=(1,\hat{q})\\
\int[d^dq]_+&=\frac{1}{4\pi}\int_0^{\infty}\frac{d\omega}{\omega^{5-d}}\int\frac{d^{d-2}\Omega_{\hat{q}}}{(2\pi)^{d-2}}
\end{align}
We take as our ordering:
\begin{align}
O(p_i,p_j;q)&=c\,\omega\Big(\frac{p_i\cdot n_q\,n_q\cdot p_j}{p_i\cdot p_j}\Big)^{\frac{\beta}{2}}
\end{align}
Within dimensional regularization, we have:
\begin{align}
\gamma_{ij}^{(1)}(\mu)&=\frac{1}{4\pi}\Big(\frac{c\bar \mu}{\mu}\Big)^{4-d}\int\frac{d^{d-2}\Omega_{\hat{q}}}{(2\pi)^{d-2}}\Bigg(\frac{p_i\cdot p_j}{p_i\cdot n_q\,n_q\cdot p_j}\Bigg)^{1-\frac{\beta}{2}(4-d)}
\end{align}
We note that when $\beta=1$, that is, when we have transverse ordering, the collinear singularities are unregulated by dimensional regularization, and require an additional regularization procedure. To cancel the collinear divergences, we must add the collinear contributions arising from jet functions, and subtract any overlap induced by the regularization procedure. For an extensive discussion, in particular in the context of virtual corrections to gauge theory amplitudes, see Ref. \cite{Chiu:2009mg}, and for renormalization/resummation of these rapidity see Refs. \cite{Becher:2010tm,Chiu:2012ir}. The end result will be to induce a maximal virtuality of the soft emission $q$ to the initial hard directions $n_i$ or $n_j$, summarized in Eq. \eqref{eq:soft_logarithm}.


\subsection{Soft Anomalous Dimension}\label{app:anom_dim}
We can consider a process by which $N$-hard partons scatter from an initial hard configuration $i$ to a final hard configuration $f$, in the presence of an external current $j$ which only couples to long wave length modes. We then have the factorization:
\begin{align}
\mathcal{A}(i\rightarrow f; j)&= \mathbf{C}_{N}(i\rightarrow f;\mu)\mathbf{Y}_N[j,\mu]\prod_{i=1}^NJ_{i}(\omega_i,\mu)\,,\\
\label{eq:anom_dim_amplitudes}\mu\frac{d}{d\mu}\mathbf{C}_{N}(i\rightarrow f;\mu)&=\mathbf{C}_{N}(i\rightarrow f;\mu)\mathbf{\Gamma}_N\Big(\{p_i\}_{i=1}^{N};\mu;\alpha_s(\mu)\Big)\,.
\end{align}
$\mathbf{\Gamma}_N$ is the soft anomalous dimension. The matrix element definition of the jet function is given in Eq. \eqref{eq:jet_amplitudes}, we have suppressed color and polarization indices for conciseness. First we present the results for the naive soft anomalous dimension up to two loop order for the soft anomalous dimension of wilson lines from Ref. \cite{Aybat:2006wq}.\footnote{For a discussion of the factorization constraints on this anomalous dimension, see Refs. \cite{Gardi:2009qi,Becher:2009qa}} The \emph{naive} transverse-ordered soft anomalous dimension is collinear divergent, which is cured by subtracting out the appropriate eikonal jet functions, or equivalently, performing the zero-bin subtraction, and adding in the jet function contribution.  We have:
\begin{align}
\mathbf{Y}_N[j,\mu]\prod_{i=1}^NJ_{i}(\omega_i,\mu)&=\mathcal{P}\text{exp}\Big(-\frac{1}{2}\int_{\mu_i^2}^{\mu^2}\frac{d\lambda^2}{\lambda^2}\mathbf{\Gamma}_N[\lambda,\alpha_s(\lambda)]\Big)\mathbf{Y}_N[j,\mu_i]\prod_{i=1}^NJ_{i}(\omega_i,\mu_i)\,,\\
\mu\frac{d}{d\mu}\Big(\mathbf{Y}_N[j,\mu]\prod_{i=1}^NJ_{i}(\omega_i,\mu)\Big)&=-\,\mathbf{\Gamma}_N\Big(\mathbf{Y}_N[j,\mu]\prod_{i=1}^NJ_{i}(\omega_i,\mu)\Big)\,,\\
\mathbf{\Gamma}_N&=-\frac{1}{2}\hat\gamma_K\Big(\alpha_s(\mu)\Big)\sum_{1\leq i<j\leq N }\mathbf{T}_i\cdot\mathbf{T}_{j}\text{ln}\frac{2p_i\cdot p_j+i0}{-\mu^2}+\sum_{i=1}^N\mathbf{T}_i^2\gamma_i\Big(\alpha_s(\mu)\Big)+...\,,\\
\hat\gamma_K\Big(\alpha_s(\mu)\Big)&=\frac{\alpha_s}{\pi}+\Big(\frac{\alpha_s}{\pi}\Big)^2\Bigg(C_A\Big(\frac{67}{36}-\frac{\pi^2}{12}\Big)-\frac{5}{9}n_fT_f\Bigg)+...\,.
\end{align}
We have only written the dipole contribution to the soft anomalous dimension, where $\hat\gamma_K$ is the cusp anomalous dimension with the leading Casimir factor scaled out. This dipole form is violated at three loops, \cite{Henn:2016jdu,Almelid:2017qju}, and the ``collinear terms'' are the terms proportional to $\mathbf{T}_i^2$, and the $...$ denote terms which violate the dipole form. The $i0$ prescription on the argument of the logarithm is more transparently written as:
\begin{align}
2p_i\cdot p_j+i0&=-|2p_i\cdot p_j|e^{-i\pi\lambda_{ij}}\,\text{ where } \lambda_{ij} = 1\text{ if both } i,j \text{ are incoming or out-going}, 0 \text{ otherwise. } 
\end{align}
We wish to examine the ``on-shell'' region of the soft anomalous dimension, so we drop the $i0$-prescription and assume all invariants are in the time-like region: $p_i\cdot p_j >0$. The correct imaginary part is restored with the Glauber contribution. Then we may represent the logarithm in the soft function as arising from the integration over the on-shell phase-space given as:
\begin{align}\label{eq:soft_logarithm}
\text{ln}\frac{2p_i\cdot p_j}{\mu^2}&=4\pi^2\int [d^4q]_+\mu\delta\Big(\mu-\sqrt{2W_{ij}^{-1}(q)}\Big)\theta\Big(\omega_i-\frac{n_j\cdot q}{n_i\cdot n_j}\Big)\theta\Big(\omega_j-\frac{n_i\cdot q}{n_i\cdot n_j}\Big)W_{ij}(q)\,,\\
p_i\cdot p_j&=\omega_i\omega_j\,n_i\cdot n_j\,.
\end{align}
We have factored the light-like momenta into their energy $\omega_i$ and a null direction $n_i=(1,\hat{n}_i)$. Formally, the naive soft-sector integral with transverse ordering is given by Eq. \eqref{eq:soft_integral_transverse_sliced}, not Eq. \eqref{eq:soft_logarithm}, which contains constraints that the soft parton cannot have too large an energy. These constraints are indeed actually realized in the jet function contributions to the on-shell component of the soft anomalous dimension. Finally, we introduce the resummation factor used throughout the text (see Eq. \eqref{eq:Anomalous_dim_gen_fun_solve}):
\begin{align}\label{eq:soft_evo_factor}
\mathcal{U}_{N}(\mu_F,\mu_I)&=\mathcal{P}\text{exp}\Bigg(-\int_{\mu_I}^{\mu_F}\frac{d\mu'}{\mu'}\mathbf{\Gamma}_{N}(\mu')\Bigg)
\end{align}

\subsection{Collinear Limit}\label{sec:collinear_limit_anom_dim}
For an all orders discussion of collinear limits in the case of time-like separations, see Ref. \cite{Dixon:2009ur}. For the on-shell contribution to the soft anomalous dimension, up to two loops, we may write:
{\small\begin{align}
\mathbf{\Gamma}_N\Big(\{p_k\}_{k=1}^{N};\mu\Big)&=\mathbf{\Gamma}_{N-1}\Big(\{p_k\}_{k=1}^{N}\Big|_{i\parallel j};\mu\Big)+\mathbf{\Gamma}_{ij}^{\mathbf{Sp}}\\
\mathbf{\Gamma}_{ij}^{\mathbf{Sp}}&=\mathbf{T}_i^2\gamma_i\big(\alpha_s(\mu)\big)+\mathbf{T}_j^2\gamma_j\big(\alpha_s(\mu)\big)-(\mathbf{T}_i+\mathbf{T}_j)^2\gamma_{i+j}\big(\alpha_s(\mu)\big)\nonumber\\
&\qquad-\frac{1}{2}\hat{\gamma}_K\big(\alpha_s(\mu)\big)\Big\{\mathbf{T}_i\cdot \mathbf{T}_j\text{ln}\frac{|s_{ij}|}{\mu^2}-\mathbf{T}_i\cdot(\mathbf{T}_i+\mathbf{T}_j)\text{ln}\frac{\omega_i}{\omega_i+\omega_j}-\mathbf{T}_j\cdot(\mathbf{T}_i+\mathbf{T}_j)\text{ln}\frac{\omega_j}{\omega_i+\omega_j}\Big\}
\end{align}}
Where $\{p_k\}_{k=1}^{N}\Big|_{i\parallel j}$ denotes replacing $p_i$ and $p_j$ by a null vector parallel to both, with an energy corresponding to the sum of the two energies $\omega_i$ and $\omega_j$. The new color generator for this combined direction is the sum of the old ones: $\mathbf{T}_i+\mathbf{T}_j$. The list then should be appropriately relabeled from $1$ to $N-1$. The splitting amplitude's anomalous dimension can also be expressed purely in terms of quadratic color generators using $2\mathbf{T}_i\cdot \mathbf{T}_j = (\mathbf{T}_i+\mathbf{T}_j)^2-\mathbf{T}_i^2-\mathbf{T}_j^2$, and thus it exponentiates simply.

\subsection{Anomalous dimensions of Soft Currents}
Finally, we can deduce the anomalous dimensions for the soft currents defined in the matching equation \eqref{eq:LSZ_reduction_matching}, up to eikonal jet function subtractions and the standard jet function contributions:
{\small\begin{align}
\mu\frac{d}{d\mu}\mathbf{J}_{N}(q_1^{\mu_1 b_1},...,q_n^{\mu_n b_n},\mu)&=\mathbf{J}_{N}(q_1^{\mu_1 b_1},...,q_n^{\mu_n b_n},\mu)\mathbf{\Gamma}_{N+n}\Big(\{p_i\}_{i=1}^{N}\cup\{\{q_i\}_{i=1}^{n}\};\mu\Big)\nonumber\\
&\qquad-\mathbf{\Gamma}_{N}\Big(\{p_i\}_{i=1}^{N};\mu\Big)\mathbf{J}_{N}(q_1^{\mu_1 b_1},...,q_n^{\mu_n b_n},\mu)\,.
\end{align}}

\bibliography{ngl_factorization}

\end{document}